\documentclass[12pt,preprint]{aastex}
\usepackage{psfig,lscape} 

\def\be{\begin{equation}}
\def\ee{\end{equation}}

\def\m{~$\mu$m}

\def\HII  {\ion{H}{2}}

\def\ISO{{\it ISO}}
\def\IRAS{{\it IRAS}}
\def\Spitzer{{\it Spitzer}}
\def\GALEX{{\it GALEX}}

\def\Spitzercolora {$f_\nu (70 \mu {\rm m})   / f_\nu (160 \mu {\rm m})$}

\begin {document}
\slugcomment{\scriptsize \today \hskip 0.2in Version 4.2}

\title{An Ultraviolet-to-Radio Broadband Spectral Atlas of Nearby Galaxies}

\author{D.A.~Dale\altaffilmark{1}, A.~Gil~de~Paz\altaffilmark{2}, K.D.~Gordon\altaffilmark{3}, H.M.~Hanson\altaffilmark{1}, L.~Armus\altaffilmark{4}, G.J.~Bendo\altaffilmark{5}, L.~Bianchi\altaffilmark{6}, M.~Block\altaffilmark{3}, S.~Boissier\altaffilmark{7,8}, A.~Boselli\altaffilmark{8}, B.A.~Buckalew\altaffilmark{9}, V.~Buat\altaffilmark{8}, D.~Burgarella\altaffilmark{8}, D.~Calzetti\altaffilmark{10}, J.M.~Cannon\altaffilmark{11}, C.W.~Engelbracht\altaffilmark{3}, G.~Helou\altaffilmark{9}, D.J.~Hollenbach\altaffilmark{12}, T.H.~Jarrett\altaffilmark{4}, R.C.~Kennicutt\altaffilmark{13,3}, C.~Leitherer\altaffilmark{10}, A.~Li\altaffilmark{14}, B.F.~Madore\altaffilmark{7}, M.J.~Meyer\altaffilmark{10}, E.J.~Murphy\altaffilmark{15}, M.W.~Regan\altaffilmark{10}, H.~Roussel\altaffilmark{16}, J.D.T.~Smith\altaffilmark{3}, M.L.~Sosey\altaffilmark{10}, D.A.~Thilker\altaffilmark{6}, F.~Walter\altaffilmark{16}}
\altaffiltext{1}{\scriptsize Department of Physics and Astronomy, University of Wyoming, Laramie, WY 82071; ddale@uwyo.edu}
\altaffiltext{2}{\scriptsize Departamento de Astrofisica, Universidad Complutense, Avenida de la Complutense s/n, Madrid, E-28040, Spain}
\altaffiltext{3}{\scriptsize Steward Observatory, University of Arizona, 933 North Cherry Avenue, Tucson, AZ 85721}
\altaffiltext{4}{\scriptsize Spitzer Science Center, California Institute of Technology, M.S. 220-6, Pasadena, CA 91125}
\altaffiltext{5}{\scriptsize Astrophysics Group, Imperial College, Blackett Laboratory, Prince Consort Road, London SW7 2AZ United Kingdom}
\altaffiltext{6}{\scriptsize Center for Astrophysical Sciences, The Johns Hopkins University, 3400 N. Charles St., Baltimore, MD 21218}
\altaffiltext{7}{\scriptsize Carnegie Observatories, Carnegie Institution of Washington, 813 Santa Barbara Street, Pasadena, CA 91101}
\altaffiltext{8}{\scriptsize Laboratoire d'Astrophysique de Marseille, B.P. 8, Traverse du Siphon, F-13376 Marseille, France}
\altaffiltext{9}{\scriptsize California Institute of Technology, MC 314-6, Pasadena, CA 91101}
\altaffiltext{10}{\scriptsize Space Telescope Science Institute, 3700 San Martin Drive, Baltimore, MD 21218}
\altaffiltext{11}{\scriptsize Astronomy Department, Wesleyan University, Middletown, CT 06459}
\altaffiltext{12}{\scriptsize NASA/Ames Research Center, MS 245-6, Moffett Field, CA 94035}
\altaffiltext{13}{\scriptsize Institute of Astronomy, University of Cambridge, Cambridge CB3 0HA, United Kingdom}
\altaffiltext{14}{\scriptsize Department of Physics and Astronomy, University of Missouri, Columbia, MO 65211}
\altaffiltext{15}{\scriptsize Department of Astronomy, Yale University, New Haven, CT 06520}
\altaffiltext{16}{\scriptsize Max Planck Institut f\"{u}r Astronomie, K\"{o}nigstuhl 17, 69117 Heidelberg, Germany}

\begin {abstract}
The ultraviolet-to-radio continuum spectral energy distributions are presented for all 75 galaxies in the \Spitzer\ Infrared Nearby Galaxies Survey (SINGS).  A principal component analysis of the sample shows that most of the sample's spectral variations stem from two underlying components, one representative of a galaxy with a low infrared-to-ultraviolet ratio and one representative of a galaxy with a high infrared-to-ultraviolet ratio.  The influence of several parameters on the infrared-to-ultraviolet ratio is studied (e.g., optical morphology, disk inclination, far-infrared color, ultraviolet spectral slope, and star formation history).  Consistent with our understanding of normal star-forming galaxies, the SINGS sample of galaxies in comparison to more actively star-forming galaxies exhibits a larger dispersion in the infrared-to-ultraviolet versus ultraviolet spectral slope correlation.  Early type galaxies, exhibiting low star formation rates and high optical surface brightnesses, have the most discrepant infrared-to-ultraviolet correlation.
These results suggest that the star formation history may be the dominant regulator of the broadband spectral variations between galaxies.  Finally, a new discovery shows that the 24\m\ morphology 
can be a useful tool for parametrizing the global dust temperature and ultraviolet extinction in nearby galaxies.  The dust emission in dwarf/irregular galaxies is clumpy and warm accompanied by low ultraviolet extinction, while in spiral galaxies there is typically a much larger diffuse component of cooler dust and average ultraviolet extinction.  
For galaxies with nuclear 24\m\ emission, the dust temperature and ultraviolet extinction are relatively high compared to disk galaxies.

\end {abstract}


\section {Introduction}
Interstellar dust has always presented challenges to astronomers.  Extinction makes it difficult to extract intrinsic fluxes.
Reddening leads to uncertain colors.  An outstanding challenge is to identify dust absorption features (diffuse interstellar bands) that were discovered over 80 years ago.  Nonetheless, interstellar dust also provides unique opportunities for understanding galaxy structure and evolution.  The formation of molecules, interstellar heating and cooling processes, polarization, and photometric redshift indicators are just a few of the areas of study that benefit from the presence and knowledge of interstellar grains (see Draine 2003 for a review).

Though dust primarily releases energy over infrared and submillimeter wavelengths, much of the radiation intercepted by interstellar grains originates in the ultraviolet from the atmospheres of OB stars.  
Thus the combination of infrared and ultraviolet data should provide a powerful diagnostic of star-formation and selective extinction.
One important application is determining ultraviolet-based star formation rates corrected for dust extinction.  High redshift surveys carried out in the rest-frame ultraviolet and optical, for example, 
are particularly vulnerable to the presence of interstellar dust (e.g., Adelberger \& Steidel 2000).  Fortunately, studies coupling infrared and ultraviolet data have shown that the slope of the ultraviolet continuum is one such useful probe of the extinction in starburst galaxies (e.g., Calzetti, Kinney, \& Storchi-Bergmann 1994; Meurer, Heckman, \& Calzetti 1999).  Subsequent work in this area has explored how the infrared-to-ultraviolet ratio and its scatter depend on bolometric and monochromatic luminosity, ultraviolet spectral slope, metallicity, diameter, star formation rate, etc. (e.g., Gordon et al. 2000; Buat et al. 2002; Bell 2003; Gordon et al. 2004; Kong et al. 2004; Buat et al. 2005; Burgarella, Buat, \& Iglesias-P\'aramo 2005; Calzetti et al. 2005; Seibert et al. 2005; Cortese et al. 2006; Schmitt et al. 2006; Iglesias-P\'aramo et al. 2006; Inoue et al. 2006).  One consistent result relevant to the work presented here is that normal star-forming (non-starburst) galaxies show larger scatter in plots of the infrared-to-ultraviolet ratio as a function of the ultraviolet spectral slope, with normal galaxies systematically exhibiting redder slopes than starburst galaxies.  This broadening in the trend has been attributed to geometry, integrated versus local extractions, and/or the increased fractional contributions from recent (versus current) star formation (e.g., Bell et al. 2002; Kong et al. 2004; Calzetti et al. 2005; Seibert et al. 2005; Boissier et al. 2006).

We are interested in exploring how the infrared-to-ultraviolet ratio depends on quantities like optical and mid-infrared morphology, ultraviolet and far-infrared color, and geometry within the SINGS sample (Kennicutt et al. 2003).  But in broader terms, the main focus of this paper is to simply present a panchromatic atlas of the broadband spectral energy distributions of a large, diverse sample of nearby galaxies, and to quantify the variety of spectral shapes evident in such a sample.  Since the fluxes presented in this work span wavelengths from the far-ultraviolet to the radio and are integrated over entire galaxies, this dataset should prove useful to astronomers studying galaxies at high redshifts, where only information on the global properties of galaxies is accessible and the rest-frame ultraviolet data are shifted into optical bandpasses.  One may plausibly argue that the variety of luminosities and spectral shapes typically seen in high redshift surveys will be different than the diversity presented below for the SINGS sample (e.g., $10^7\lesssim L_{\rm IR}/L_\odot \lesssim10^{11}$), since flux-limited surveys at high redshifts will mainly be sampling luminous and infrared-warm systems.  On the other hand, deep far-infrared surveys show significant numbers of higher redshift systems similar to local normal star-forming galaxies in mass, size, and dust temperature (e.g., Chapman et al. 2002; Sajina et al. 2006).  In either case, the rich collection of {\it Spitzer}, GALEX, and ancillary data provided by the SINGS project represents an important panchromatic baseline for extragalactic work.  

Some of the analysis presented below could be accomplished using existing datasets, for example the GALEX$+${\it UBV}$+$2MASS+{\it IRAS} work of Gil de Paz et al. (2006).  However, the sensitivity and angular resolution of our {\it Spitzer} observations allow us to probe the dust emission in both bright and faint galaxies, and to do so in a spatially resolved manner.  The paper is outlined as follows.  Section~\ref{sec:sample} presents the SINGS sample while Section~\ref{sec:data} presents the collection of ultraviolet, optical, near-infrared, infrared, submillimeter, and radio data.  The analysis of the broadband spectral energy distributions is described in Section~\ref{sec:seds} and the infrared-to-ultraviolet ratio is explored in detail in Section~\ref{sec:uvir}.  A discussion and summary of the main results are provided in Section~\ref{sec:summary}.

\section {The Sample}
\label{sec:sample}
The selection of the 75 galaxies in the \Spitzer\ Nearby Galaxies Survey (SINGS; Kennicutt et al. 2003) aimed to span a wide range in three key parameters (optical morphology, luminosity, infrared-to-optical ratio) and to adequately sample several other secondary parameters (e.g., infrared color, metallicity, surface brightness, inclination, bar structure, etc.).  The SINGS sample is comprised of nearby galaxies, with a median distance of $\sim$10~Mpc and a maximum distance of 30~Mpc.  SINGS galaxies come from a wide range of environments and galaxy types: low-metallicity dwarfs; quiescent ellipticals; dusty grand design spirals; Seyferts, LINERs, and star-forming nuclei of normal galaxies; systems within the Local and M~81 groups; and both field and (Virgo) cluster galaxies (Table~\ref{tab:additional_data}).  

\section {The Data}
\label{sec:data}

Tables~\ref{tab:uv_opt}-\ref{tab:infrared} present the global flux densities for the entire SINGS sample, for wavelengths spanning the ultraviolet through the radio.  The data are corrected for Galactic extinction (Schlegel, Finkbeiner, \& Davis 1998) assuming $A_V/E(B-V)\approx3.1$ and the reddening curve of Li \& Draine (2001).  The effect of airmass has been removed from the ground-based fluxes.  Below follows a description of the new ultraviolet and optical and archival radio data collected for the SINGS program, in addition to a few updates to the $Spitzer$ data presented in Dale et al. (2005).  

\subsection {Ultraviolet Data}

The GALEX mission (Martin et al. 2005) is performing an all-sky survey at ultraviolet wavelengths.  The imaging portion of the survey is being carried out with a far-ultraviolet and a near-ultraviolet filter centered at 1528 and 2271~\AA\ with respective full-widths at half-maximum of 269 and 616\AA\ FHWM.  In addition to imaging the entire sky with an effective exposure time of $\sim$0.1~ksec, GALEX is also carrying out relatively deep integrations ($\sim$1.5~ksec) for a few hundred nearby galaxies, including nearly the entire SINGS sample.
With an angular resolution of 4-6\arcsec, the spatial details in GALEX images are well matched to those seen in \Spitzer\ 24\m\ imaging and more resolved than in \Spitzer\ 70 and 160\m\ images.  At the median distance of the SINGS sample ($\sim$10~Mpc), the GALEX and MIPS 24\m\ data probe spatial scales of about $\sim$300~pc.  This resolution coupled with the GALEX field-of-view of 1\fdg25 allow for robust measures of sky-subtracted, spatially-integrated ultraviolet fluxes even for large nearby galaxies.

Integrated ultraviolet fluxes are computed from the surface photometry profiles derived for the {\it GALEX Atlas of Nearby Galaxies} (Gil de Paz et al. 2006).\footnote{A few SINGS sources are not in the {\it GALEX Atlas of Nearby Galaxies}, but the observing and data reduction procedures for these galaxies are the same as for the {\it Atlas} targets (e.g., M81~Dwarf~A, NGC~3773, NGC~4254, NGC~4725, NGC~6882, and NGC~6946).}  
Table~\ref{tab:uv_opt} lists the global fluxes that include an asymptotic extrapolation to the isophotal profiles.  The extrapolations are typically small and result in asymptotic fluxes that are, on average, 14\% larger than those obtained at the optical radius; $\left< f_{\rm UV}({\rm asymptotic}) / f_{\rm UV}(R_{25}) \right> = 1.14$ with a dispersion of 0.16 and 0.14 in the far- and near-ultraviolet, respectively.  Foreground field stars and background galaxies were masked before flux extraction (see Gil de Paz et al. 2006).
 Some of the SINGS galaxies have not yet been observed with GALEX but observations are soon planned (NGC~1377, NGC~3184, NGC~5033, and IC~4710), and a few only have near-ultraviolet observations because the far-ultraviolet detector was turned off at that time (see Table~\ref{tab:uv_opt}).
Bright nearby stars make it unlikely GALEX will obtain data for NGC~5408.  

The uncertainties listed in Table~\ref{tab:uv_opt} include the formal uncertainties from the weighted fits to the growth curves using the uncertainties of the individual points in the growth curves, in addition to absolute calibration uncertainties of $\sim$15\% in both the far- and near-ultraviolet. 

The average far-ultraviolet radiation field can be estimated from the far-ultraviolet fluxes and a beam size that characterizes the area from which the far-ultraviolet flux is emitted.  In units of the local Milky Way field ($1.6 \cdot 10^{-3}$~erg~cm$^{-2}$~s$^{-1}$), the average far-ultraviolet radiation field can be expressed as
\be
G_0 = 2.2 \cdot 10^3 {\left[ f_\nu(1528\AA) \over {\rm mJy}\right]} \left[ {{\rm arcsec} \over \theta_{\rm FUV}} \right]^2,
\ee
where $\theta_{\rm FUV}$ is the equivalent radius of the ellipse including half of the total far-ultraviolet light (see Tables~\ref{tab:additional_data} and \ref{tab:uv_opt}).  Typical values for the SINGS sample span $1\lesssim G_0 \lesssim 25$ with a median value of $G_0=7.4$.

\subsection {Optical Data}

Though RC3 fluxes (de Vaucouleurs et al. 1991) in the $B$ and $V$ bands are available for a large portion of the SINGS sample, we pursued a {\it BVRI} imaging campaign for reasons of consistency, sensitivity, and completeness.  The optical imaging for the SINGS project was carried out over the course of five observing runs at the Kitt Peak National Observatory 2.1~m and one observing run at the Cerro Tololo Inter-American Observatory 1.5~m telescopes between March 2001 and February 2003.  Broadband photometry was obtained in (Harris) {\it BVRI} using 2K$\times$2K CCDs with pixel scales and fields-of-view of 0\farcs305 and 10\arcmin\ at KPNO and 0\farcs433 and 14\farcm5 at CTIO.  Galaxies more extended than the CCD fields-of-view were imaged at multiple, overlapping pointings.  Typical exposure times were 1440~s ($B$), 720~s ($V$), 420~s ($R$), and 840~s ($I$), usually split into two separate exposures to aid cosmic ray removal.  Such exposures reach a depth of about 25~mag~arcsec$^{-2}$ at a signal-to-noise ratio of $\sim$10 per resolution element.

Data processing consisted of standard routines such as bias subtraction,
flat-fielding with both dome- and twilight-flats, cosmic ray removal, and the mosaicking of overlapping pointings for galaxies with large angular extents.  The southern 3\arcmin\ of the KPNO 2.1~m CCD field-of-view suffers from vignetting; care is taken to remove as much of the vignetted portion of the KPNO images as feasible.  Photometric standard stars were observed during each observing run to flux calibrate the images.  The images have photometric accuracy of 5\% or better.

Global optical fluxes are extracted using the same apertures used for the IRAC and MIPS global flux extractions; these apertures cover at least the entire optical disk (see Table~\ref{tab:additional_data}) and are chosen to be large enough to encompass all of the optical and infrared emission; in many instances the extended 160\m\ emission drives the final choice of aperture.  Sky estimation and subtraction is carried out through the use of multiple sky apertures placed near the source without overlapping the faintest isophotes visible from the galaxy.  Foreground stars are edited from the optical images after first being conservatively identified using $f_\nu$(3.6\m)/$f_\nu$(8.0\m) and $f_\nu$(8.0\m)/$f_\nu$(24\m) color images (e.g., $f_\nu$(8.0\m)/$f_\nu$(24\m)$>8$ for stars).

\subsection{Infrared Data}
\label{sec:infrared}
A full description of the infrared (2MASS, $ISO$, $IRAS$, $Spitzer$) and submillimeter (SCUBA) data can be found in Dale et al. (2005).  In this section we present details of a few additional modifications and updates to the $Spitzer$ data.  For example, the MIPS flux calibrations and their uncertainties have been altered since Dale et al. (2005)---the 24, 70, and 160\m\ calibration factors have been respectively boosted by factors 1.018, 1.107, and 1.049, and their systematic uncertainties have dropped to 4, 7, and 12\% (Engelbracht et al. in prep.; Gordon et al. in prep.; Stansberry et al. in prep.).  The uncertainties provided in Table~\ref{tab:infrared} include both calibration and statistical uncertainties.  Calibration uncertainties are 5-10\% for IRAC 3.6 and 4.5\m\ data, and 10-15\% for IRAC 5.8 and 8.0\m\ data; 10\% calibration uncertainties are used in Table~\ref{tab:infrared}.

The IRAC flux densities in Table~\ref{tab:infrared} include extended source aperture corrections provided by the Spitzer Science Center\footnote{See spider.ipac.caltech.edu/staff/jarrett/irac/}.  These corrections account for the ``extended'' emission due to the wings of the PSF and also for the scattering of the diffuse emisssion across the IRAC focal plane.  For an effective aperture radius $r=\sqrt{ab}$ in arcseconds derived from the semi-major $a$ and semi-minor $b$ ellipse axes provided in Table~\ref{tab:additional_data}, the IRAC extended source aperture correction is
\be
f^{\rm IRAC}_{\rm true} / f^{\rm IRAC}_{\rm measured} = A {\rm e}^{-r^B} + C,
\label{eqn:irac}
\ee
where $A$, $B,$ and $C$ are listed in Table~\ref{tab:IRAC_ap}.  The average extended source aperture corrections ($\sim$10\% uncertain) for the SINGS IRAC photometry are [0.912,0.942,0.805,0.749] at [3.6,4.5,5.8,8.0]($\micron$).

The MIPS flux densities in Table~\ref{tab:infrared} also include extended source aperture corrections.  Three high-resolution models of a galaxy's structure are generated by convolving an $R$ band image of the galaxy with the MIPS PSFs.  The MIPS aperture corrections listed in Table~\ref{tab:submm_radio} are computed for the same apertures used in the global flux extractions (Table~\ref{tab:additional_data}).  The median aperture corrections are [1.01,1.04,1.10] at [24,70,160]($\micron$).  The uncertainties in the aperture corrections are typically a few percent, and are based on the differences between the canonical and ``minimum'' corrections.  Minimum MIPS aperture corrections are computed assuming point source light distributions.

Finally, a correction for 70\m\ non-linearity effects is included in this presentation.  A preliminary correction of the form
\be
f^{\rm 70{\scriptsize \micron}}_{\rm true} = 0.581 (f^{\rm 70{\scriptsize \micron}}_{\rm measured})^{1.13},
\ee
derived from data presented by Gordon et al. (2006, in preparation), is applied to pixel values above a threshold of $\sim$66~MJy~sr$^{-1}$.  A small fraction of the pixels in a total of 40 SINGS 70\m\ images require such a correction.  The median correction to the global 70\m\ flux density for these 40 galaxies is a factor of 1.03, with the three largest corrections being factors of 1.124 (NGC~4826), 1.128 (NGC~1482), and 1.158 (NGC~7552).

\subsection {Radio Data} 

Global 20~cm continuum fluxes from the literature are available for 62 SINGS galaxies, with data for 51 of these galaxies taken from the New VLA Sky Survey catalog
(Condon et al. 1998; Yun, Reddy, \& Condon 2001; see Table~\ref{tab:infrared}).  Although this is a snapshot survey and prone to miss extended emission from galaxies having large angular extents, proper attention has been paid to these effects to derive unbiased 1.4~GHz fluxes (e.g., Yun, Reddy, \& Condon 2001).  The 20~cm data for 11 additional galaxies were taken from Condon (1987), Hummel (1980), Condon et al. (1990), Wright \& Otrupcek (1990), Bauer et al. (2000), and Cannon et al. (2006b).


\section {Results}
\label{sec:seds}

\subsection {Global Broadband Spectral Energy Distributions}

Figures~\ref{fig:seds1}-\ref{fig:seds8} show the ultraviolet-to-submillimeter spectral energy distributions for the SINGS sample.  The solid curve is the sum of a dust (dashed) and a stellar (dotted) model.  The dust curve is a Dale \& Helou (2002) model (least squares) fitted to ratios of the 24, 70, and 160\m\ fluxes (a dust curve for NGC~3034 is fit using $IRAS$ 25, 60, and 100\m\ data, since the MIPS data for this galaxy are saturated).  The $\alpha_{\rm SED}$ listed within each panel parametrizes the distribution of dust mass as a function of heating intensity, as described in 
Dale \& Helou (2002).  
Broadly speaking, smaller values of $\alpha_{\rm SED}$ (more heating from stronger interstellar radiation fields) correspond to later Hubble types and larger infrared-to-ultraviolet ratios.
The stellar curve is a 1~Gyr continuous star formation, solar metallicity curve from Vazquez \& Leitherer (2005) fitted to the 2MASS data.  The initial mass function for this curve utilizes a double power law form, with $\alpha_{\rm 1,IMF}=1.3$ for $0.1<m/M_\odot<0.5$ and $\alpha_{\rm 2,IMF}=2.3$ for $0.5<m/M_\odot<100$ (e.g., Kroupa 2002).  Though this stellar curve (not adjusted for internal extinction) may not be applicable to many galaxies, especially ellipticals, it is included as a ``standard'' reference against which deviations in the ultraviolet and optical can be compared from galaxy to galaxy.  The stellar curve also serves to highlight the relative importance of stars and dust in each galaxy, particularly in the transition from stellar to dust emission in the mid-infrared (e.g. NGC~1404 versus NGC~1482).

Several galaxies show mid-infrared data that deviate from the fits.  Most of these systems are low metallicity objects (e.g., Ho~II, NGC~2915, IC~2574, DDO~154, DDO~165, and NGC~6822), objects that have been shown to be deficient in PAH emission (see the discussion in Section~\ref{sec:binned}).  The mid-infrared data for NGC~1377 are also quite discrepant from the model, showing a strong excess for each of the broadband filters from 3.6 to 15\m.  The substantial hot dust emission and lack of optical signatures or synchrotron radiation led Roussel et al. (2003) to infer that this heavily extincted system is undergoing the very beginnings of an intense burst of star formation.


\subsection {Spectral Energy Distributions Binned by the Infrared-to-Ultraviolet Ratio}
\label{sec:binned}

\begin{figure}
 \plotone{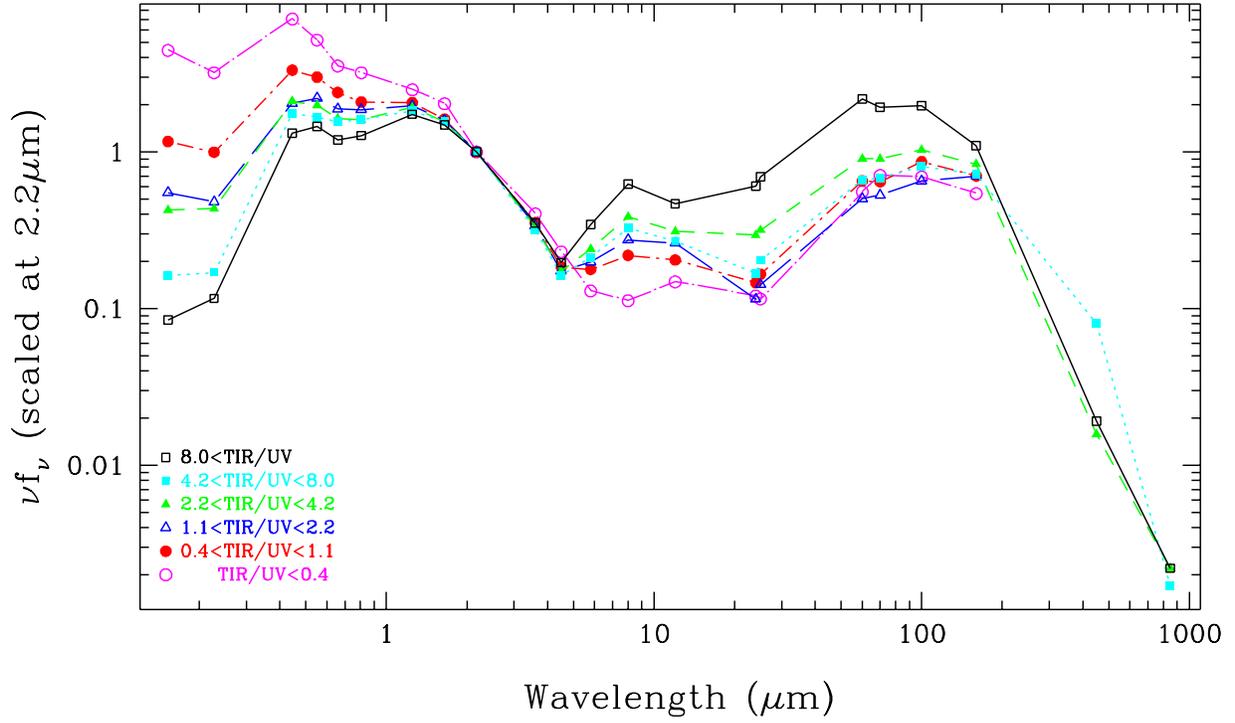}
 \caption{A display of stacked spectral energy distributions that emphasizes the infrared-to-ultraviolet variations within the SINGS sample.  Each spectral energy distribution in the stack represents an average of approximately 10 individual spectral energy distributions that fall within a given bin of the infrared-to-ultraviolet ratio.}
 \label{fig:stack}
\end{figure}

Analysis of the distribution of global (spatially-integrated) spectral energy distributions is a sensible starting point for current cosmology surveys (e.g., Rowan-Robinson et al. 2005).  Figure~\ref{fig:stack} shows a stack of SINGS spectral energy distributions that emphasizes the infrared-to-ultraviolet variations within the SINGS sample.  Each spectral energy distribution in the stack represents an average of approximately 10 individual spectral energy distributions that fall within a given bin of the total infrared-to-ultraviolet ratio (TIR/[FUV+NUV]; the bins are indicated in the figure legend).  The ultraviolet emission for this ratio is computed as $\nu f_\nu$(1500\AA)+$\nu f_\nu$(2300\AA) whereas the ``total infrared'' is the dust continuum emission between 3 and 1100\m\ (Dale et al. 2001), computed using the MIPS 24, 70, and 160\m\ fluxes and Equation~4 of Dale \& Helou (2002).  The spectra are arbitrarily normalized at the 2MASS $K_{\rm s}$ band wavelength.  

Several features in the stack are immediately noticeable.  The ultraviolet slopes vary from positive values for galaxies with high infrared-to-ultraviolet ratios to negative values for low infrared-to-ultraviolet ratio galaxies (as will be explored in detail in \S~\ref{sec:beta}).  The 4000\AA\ break shows up quite clearly, even at this coarse spectral ``resolution.''  Other obvious features include: the broad far-infrared peak signifying emission from cool-to-warm large grains; the contributions from polycyclic aromatic hydrocarbons appearing as mid-infrared emission features; the near-infrared hump arising from photospheric emission from old stellar populations; and a near-infrared H$-$ opacity signature for high infrared-to-ultraviolet systems.  Note also the broad spread in the ultraviolet data compared to that in the far-infrared.  Since the spectra are normalized at $K_{\rm s}$, this difference implies that there is a larger spread in ultraviolet light per unit stellar mass than infrared light per unit stellar mass (Gil de Paz et al. 2006, for example, find that [{\it FUV}$-${\it K}] spans 11 mag in their GALEX {\it Atlas}).  The variations in the infrared-to-ultraviolet ratio studied later in this work are largely driven by variations in the ultraviolet emission.

Close inspection of Figure~\ref{fig:stack} reveals that most of the variation in the stacked spectra stem from the two extreme bins (bins ``1'' and ``6'') in the infrared-to-ultraviolet ratio.  However, substantial variations are still seen in bins 2-5 at ultraviolet and mid-infrared wavelengths.  The bin 2-5 range is 0.88, 0.78, 0.24, and 0.16~dex at 0.15, 0.23, 8.0, and 24\m\ (compared to the bin 1-6 range of 1.76, 1.46, 0.80, and 0.80~dex at the same wavelengths).  The spread at ultraviolet wavelengths is presumably significantly affected by variations in dust extinction.  The range in 8.0\m\ emission, on the other hand, is likely due to PAH destruction/formation variations.  Low metallicity systems, for example, are known to be deficient in PAH emission (e.g., Dale et al. 2005; Engelbracht et al. 2005; Galliano et al. 2005; Walter et al. 2006).  Indeed, eight of the nine galaxies in the lowest infrared-to-ultraviolet ratio bin have low metallicities ($12+\log({\rm O/H})<8.1$; Moustakas et al. 2006, in preparation), and this bin's average spectrum in Figure~\ref{fig:stack} shows very low mid-infrared emission.  The 24\m\ emission from galaxies is known to be sensitive to the star formation rate (e.g., Dale et al. 2005; Gordon et al. 2004; Helou et al. 2004; Hinz et al. 2004; Calzetti et al. 2005); the observed variations at this wavelength may be strongly affected by the range in the sample's star formation properties.





\subsection {Principal Component Analysis}
\label{sec:pca}

A principal component analysis can help to quantify relative contributions to the observed variations in a sample of spectral energy distributions (Deeming 1964).  A set of $i$ eigenvectors $\{\vec{e}_i\}$ and their corresponding eigenvalues $\{e_i\}$ for our sample of $N$ galaxies are computed from a diagonalization of the covariance matrix
\begin{equation}
C_{jk}=\frac{1}{N}\Sigma_{i=1}^{N} \; \nu f^i_{\nu}(\lambda_j) \; \nu f^i_{\nu}(\lambda_k),
\end{equation}
where $\nu f^i_{\nu}(\lambda_j)$ is the flux of the $i^{\rm th}$ spectrum at wavelength $\lambda_j$.  We restrict the computation of the covariance matrix to involve only those wavelengths for which we have a substantial database of fluxes; submillimeter data at 450 and 850\m\ are not included in the principal component analysis.  Furthermore, to avoid spurious results we do not include in our analysis any SINGS galaxies without a secure detection/measurement at any of the ultraviolet, optical, near-infrared, or infrared wavelengths listed in Tables~\ref{tab:uv_opt}-\ref{tab:infrared}.  Hence, our principal component analysis involves only about three-fourths of the SINGS sample (Table~\ref{tab:additional_data} indicates which systems are involved).
Our principal component analysis is carried out after normalizing the spectra 
at the 2MASS $K_{\rm s}$ band wavelength.  

\begin{figure}
 \plotone{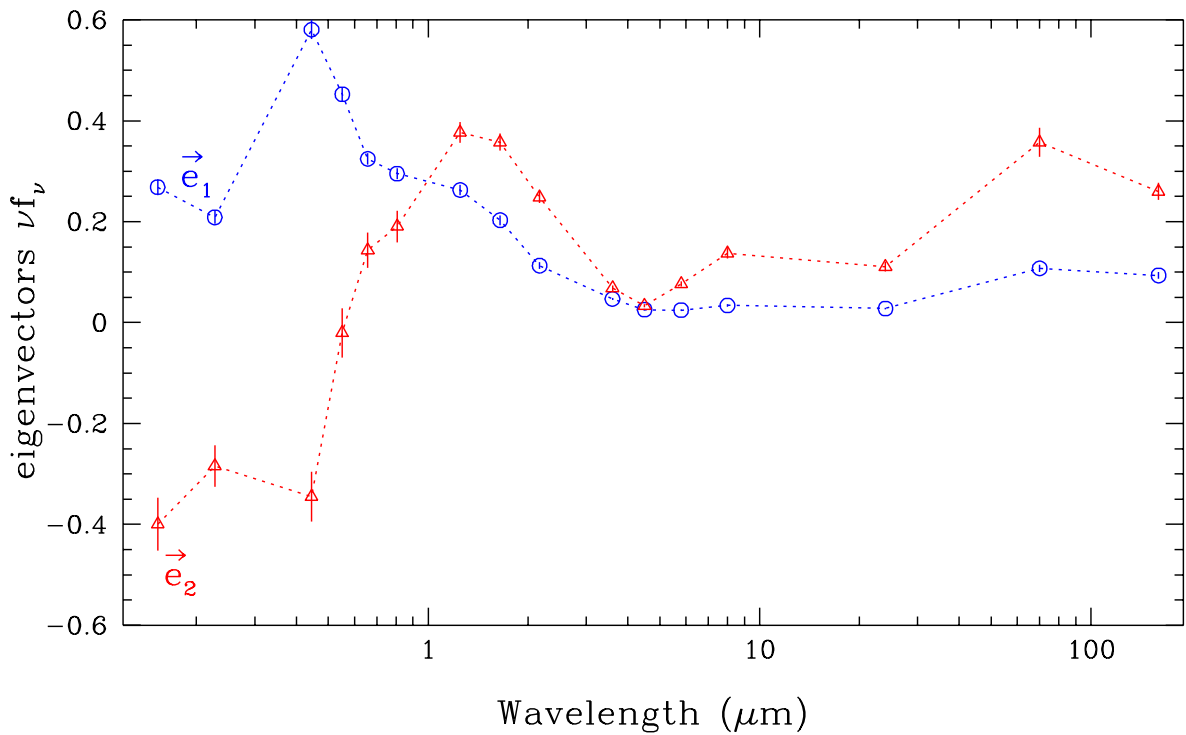}
 \caption{The strongest (circles) and second strongest (triangles) eigenvector spectra from a principal component analysis of the SINGS spectra are displayed.  These are average eigenvectors stemming from 10,000 Monte Carlo simulations based on the observed fluxes and their uncertainties (corrected for Galactic extinction and airmass in the case of ground-based observations); the error bars shown in this figure indicate the dispersion of the eigenspectra from the simulations.  These eigenvectors have normalized eigenvalues of 0.88 and 0.07; $\left< \vec{e}_1 \right>$ and $\left< \vec{e}_2 \right>$ respectively contribute to 88\% and 7\% of the observed variation in the sample spectra.}
 \label{fig:eigenvectors}
\end{figure}

The principal component analysis
has produced eigenvectors that describe different components of the 
sample spectra (Figure~\ref{fig:eigenvectors}).
The most important eigenvectors are the first two.  Eigenvector $\vec{e_1}$ appears to describe the contribution of unobscured starlight (including starlight from star formation regions) to the galaxies' spectra.  This eigenvector also includes additional low-level far-infrared emission that may represent cirrus emission from the diffuse interstellar medium.  Eigenvector $\vec{e_2}$ appears to describe the effects of dust on the shape of the spectrum.  This eigenvector demonstrates that as the dust content of the galaxies increases, the mid- and far-infrared flux densities increase while the starlight becomes redder (i.e., the ultraviolet and blue light decreases while the red and near-infrared light increases).

The individual eigenvalues normalized by the sum $\Sigma_j e_j$ of all eigenvalues indicates the fractional contribution of each corresponding eigenvector to the variation in the spectral atlas (when normalized at $K_{\rm s}$).  The normalized $e_1$ and $e_2$ eigenvalues (which correspond to the $\vec{e_1}$ and $\vec{e_2}$ eigenvectors) are 0.88 and 0.07, respectively.  In other words, $\vec{e_1}$ and $\vec{e_2}$ represent 88\% and 7\% of the total variation in the spectral atlas.  The remaining 5\% of the variation is represented by additional eigenvectors with corresponding normalized eigenvalues that are individually 0.02 or smaller.  To quantify the uncertainty on these numbers, we have performed 10,000 Monte Carlo simulations of the principal component analysis.  For each simulation we use the tabulated flux uncertainties to add a random (Gaussian deviate) flux offset to every galaxy's flux at each wavelength.  The means of the two largest normalized eigenvalues from these simulations are $\left< e_1 \right>=0.88\pm0.01$ and $\left< e_2 \right>=0.07\pm0.01$, with the error bars reflecting the standard deviations in the simulations.  The eigenspectra and error bars in Figure~\ref{fig:eigenvectors} reflect the mean values and standard deviations in the simulations.





\section {The Infrared-to-Ultraviolet Ratio}
\label{sec:uvir}

The infrared-to-ultraviolet ratio is a rough measure of the amount of extinction at ultraviolet wavelengths.  The infrared-to-ultraviolet ratio in galaxies is also sensitive to the metal content, luminosity, star formation history, and the relative distribution of interstellar grains with respect to their heating sources.  What is the predominant driver of the variations in 
this ratio in galaxies?
Which parameters can be used to most easily quantify these variations, with the aim of simplifying SED analysis?  Various possibilities are presented and discussed below.

\subsection {Inclination}

\begin{figure}
 \plotone{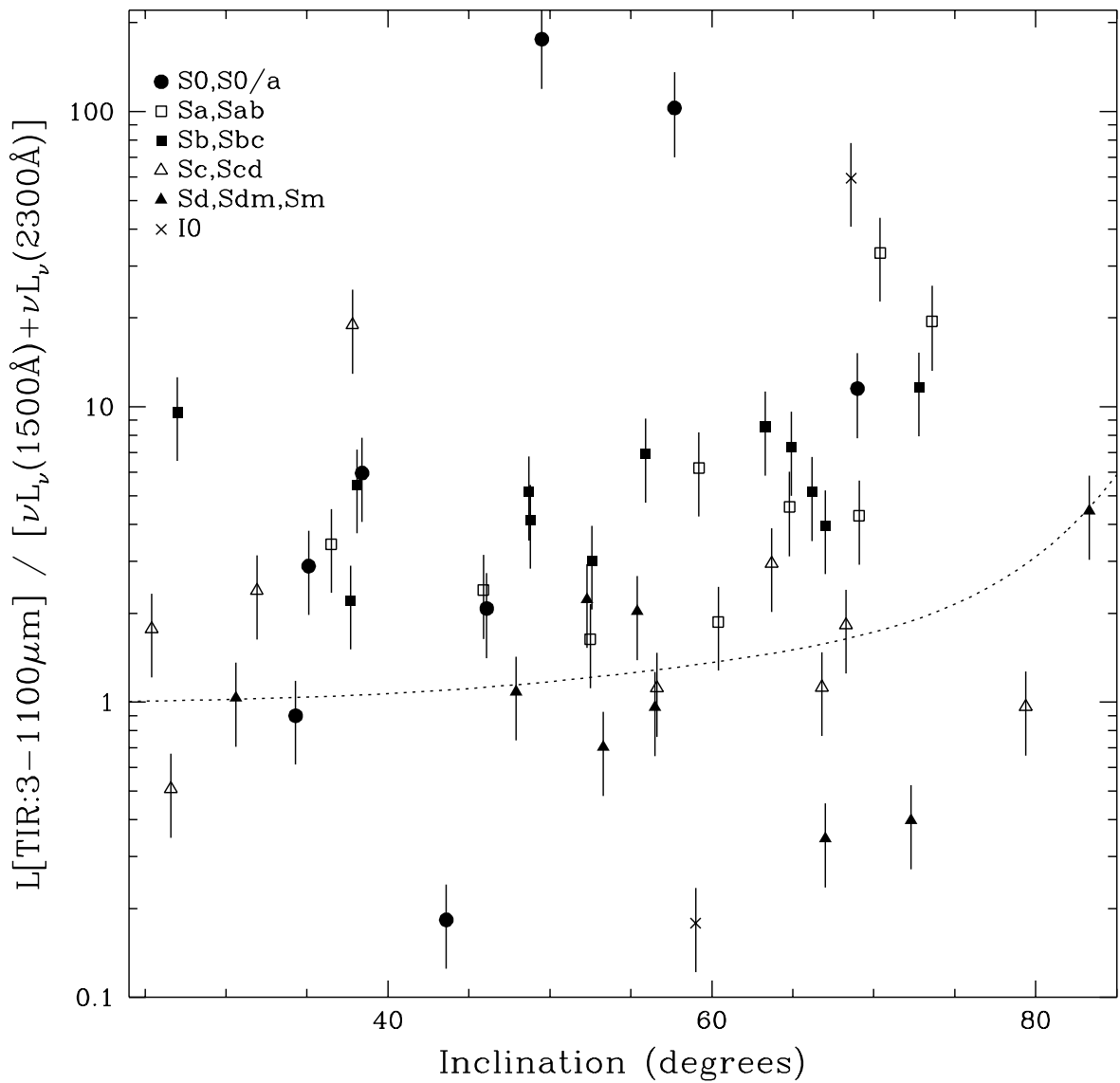}
 \caption{The infrared-to-ultraviolet ratio as a function of galaxy disk inclination.  The dotted line, normalized to an infrared-to-ultraviolet ratio of unity at zero inclination, shows the expected effect of extinction on the ultraviolet data with changing inclination using the thin disk model and a central face-on optical depth in the $B$ band of $\tau_B^{\rm f}=2$ described in Tuffs et al. (2004).  The error bars stem from the observational uncertainties.}
 \label{fig:inclination}
\end{figure}

The tilt of a spiral disk with respect to the observer's line-of-sight affects the observed intensity and colors (e.g., Bruzual, Magris, \& Calvet 1988; Boselli \& Gavazzi 1994; Giovanelli et al. 1995; Kuchinski et al. 1998).  The ``disk'' inclination can be computed from the observed semi-major and semi-minor axes, $a$ and $b$, assuming that disks are oblate spheroids with intrinsic axial ratio $(b/a)_{\rm int}$ using the relation:
\be
\cos^2 i = { (b/a)^2 - (b/a)_{\rm int}^2 \over 1-(b/a)_{\rm int}^2 },
\ee
where $(b/a)_{\rm int}\simeq0.2$ for morphological types earlier than Sbc and $(b/a)_{\rm int}\simeq0.13$ otherwise (see Dale et al. 1997 and references therein).  Figure~\ref{fig:inclination} gives the infrared-to-ultraviolet ratio as a function of galaxy disk inclination.  Galaxies with elliptical and irregular morphologies have not been included in the plot.   The dotted line (normalized to an infrared-to-ultraviolet ratio of unity at zero inclination) shows the expected effect of extinction on the ultraviolet data with changing inclination using the thin disk model and a central face-on optical depth in the $B$ band of $\tau_B^{\rm f}=2$ described in Tuffs et al. (2004).  The ratio does not obviously trend with galaxy orientation; if there is a trend consistent with the model of Tuffs et al., it is a weak trend that is washed out by a large dispersion.  The data in Figure~\ref{fig:inclination} indicate that moderate disk inclinations are not a dominant factor in determining the infrared-to-ultraviolet ratio in SINGS galaxies.

\subsection {Hubble Type}

\begin{figure}
 \plotone{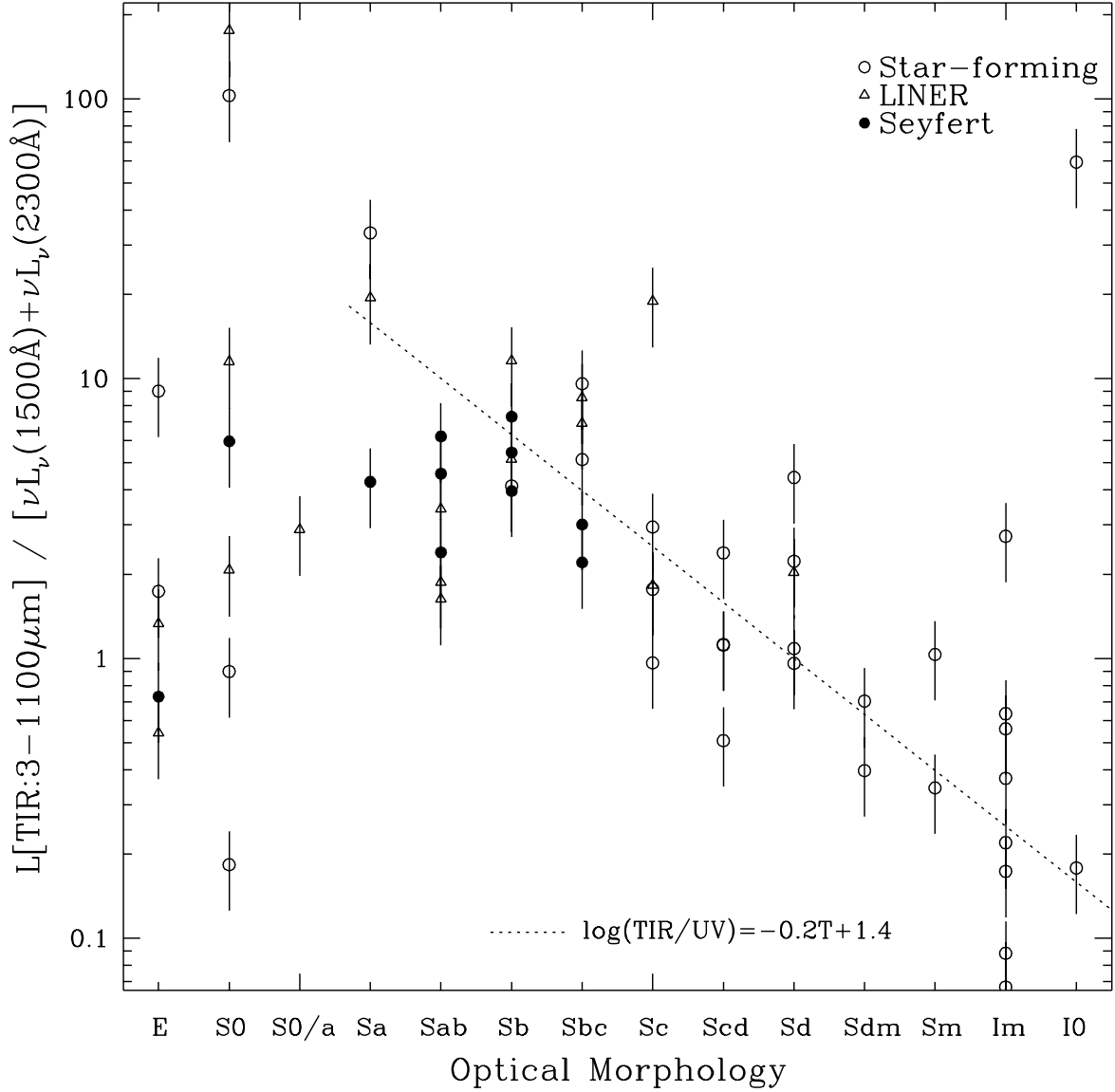}
 \caption{The infrared-to-ultraviolet ratio as a function of galaxy optical morphology.  The equation provided quantifies the approximate trend with Hubble type for late-type galaxies shown as a dotted curve (e.g., Sa$\rightarrow T=$1, Sb$\rightarrow T=$3, Sc$\rightarrow T=$5, etc.).}
 \label{fig:morphology}
\end{figure}

Figure~\ref{fig:morphology} displays the infrared-to-ultraviolet ratio as a function of galaxy optical morphology.  In general, the ultraviolet light increases in importance and the dust emission decreases in importance as the morphology changes from early-type spirals to late-type spirals to irregulars, reflecting the changing significance of recent star formation and 
dust content/distribution
to the overall energy budget in galaxies (see also Buat \& Xu 1996 and Gil de Paz et al. 2006).  One interpretation is that the infrared-to-ultraviolet ratio increases as the redder, older stellar populations increasingly dominates for earlier-type spirals, but as will be shown in \S~\ref{sec:geometry}, the increased porosity of star-forming regions for more actively star-forming galaxies may play a key role in this trend.  Elliptical and S0 galaxies do not follow the general trend exhibited by the spirals and irregulars; some ellipticals and S0s show comparatively low infrared-to-ultraviolet ratios.  
This deviation to low infrared-to-ultraviolet ratios for some of the earliest-type galaxies 
could be due to a relative paucity of dust grains intercepting ultraviolet/optical/near-infrared photons and reprocessing that energy into the infrared; the infrared portion of the bolometric luminosity in ellipticals is typically only a few percent (Xilouris et al. 2004).  Moreover, some elliptical systems are conspicuous ultraviolet emitters, with the emission thought to mainly arise from low-mass, helium-burning stars from the extreme horizontal branch and later phases of stellar evolution (see O'Connell 1999 for a review).  Low or moderate levels of star formation could also contribute to the ultraviolet emission in early-type galaxies (e.g., Fukugita et al. 2004).  Recent evidence shows that strong ultraviolet emitters are the largest contributers to the significant scatter in the ultraviolet colors of early-type galaxies (e.g., Yi et al. 2005; Rich et al. 2005).  



This wide range in the fractional ultraviolet luminosity also leads to significant scatter in the infrared-to-ultraviolet ratio.  Though the statistics are based on small numbers, a similarly large dispersion is seen for irregular systems at the other end of the morphological spectrum.  Part of this dispersion is likely associated with the metal content in irregular/dwarf systems.  In general, irregular galaxies are quite blue and metal-poor (e.g., Hunter \& Gallagher 1986; van Zee, Haynes, \& Salzer 1997).  Ultraviolet/optical continuum emission from low-metallicity galaxies experiences less extinction, which inhibits the production of infrared continuum emission (see previous paragraph).  The combination of these effects leads to lower infrared-to-ultraviolet ratios in irregular galaxies.\footnote{The amorphous, starbursting M~82 hosts a large infrared-to-ultraviolet ratio and should be considered separately from the dwarf irregulars.}



\subsection {Far-Infrared Color}
\label{sec:geometry}

\begin{figure}
 \plotone{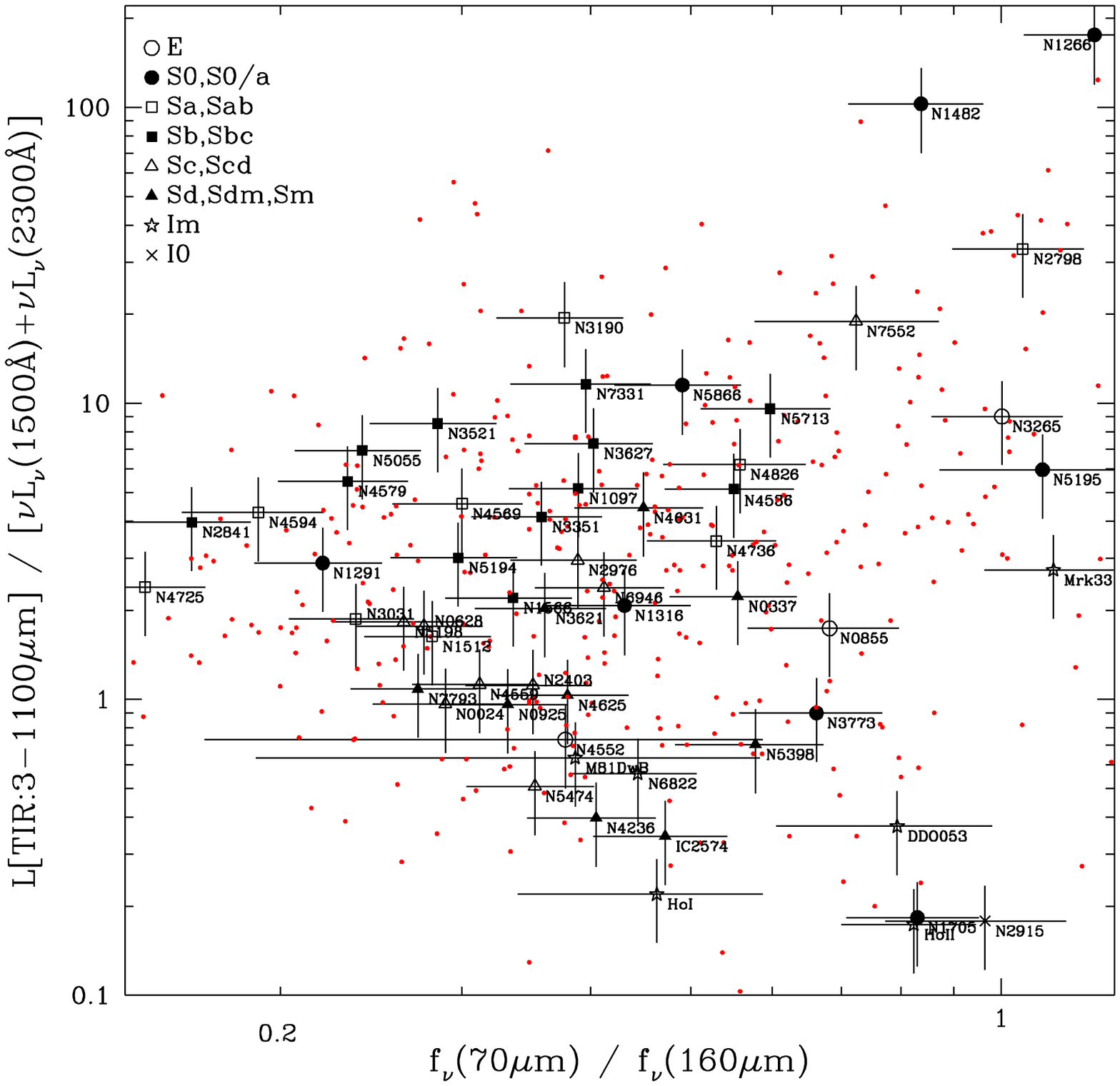}
 \caption{The infrared-to-ultraviolet ratio as a function of far-infrared color for the SINGS sample.  Data from the {\it GALEX Atlas of Nearby Galaxies} (Gil de Paz et al. 2006) are also shown as small data points without error bars.}
 \label{fig:ircolor}
\end{figure}

Though dwarf irregulars show low infrared-to-ultraviolet ratios, their interstellar dust grains tend to be vigorously heated.  The lower metallicity in these systems results in less line blanketing which in turn leads to harder radiation fields.  Many of the dwarf and irregular systems in the SINGS sample indeed have elevated $f_\nu(70\micron)/f_\nu(160\micron)$ ratios (e.g., Dale et al. 2005; Walter et al. 2006; see also Boselli, Gavazzi, \& Sanvito 2003), indicating strong overall heating of the dust grain population.  The warmer far-infrared colors for SINGS dwarfs/irregulars are evident in Figure~\ref{fig:ircolor}, which plots the infrared-to-ultraviolet ratio versus far-infrared color for the entire SINGS sample.  

An interesting feature to this plot is the apparent wedge-shaped distribution, with a progressively smaller range in the infrared-to-ultraviolet ratio for cooler far-infrared colors.  There is no obvious trend in infrared-to-ultraviolet ratio with disk inclination (Figure~\ref{fig:inclination}), so it is unlikely that the distribution in Figure~\ref{fig:ircolor} is due solely to disk orientation.  
The small data points without error bars come from the ({\it IRAS}-based) {\it GALEX Atlas of Nearby Galaxies} (Gil de Paz et al. 2006) and follow the same general distribution as the SINGS data, suggesting that this wedge-shaped distribution is unlikely a sample selection effect. 
Presumably the upper left-hand portion of this figure, for example, is empty since a large infrared-to-ultraviolet ratio requires lots of dust opacity, 
but higher opacity implies a larger density of interstellar dust closer to heating sources, therefore leading to warm dust and high values of $f_\nu(70\micron)/f_\nu(160\micron)$.

The relative distribution of dust grains and their heating sources may, in fact, play a key role in creating this overall wedge-shaped distribution.  As argued above, it is reasonable to assume that galaxies with relatively high $f_\nu(70\micron)/f_\nu(160\micron)$ ratios have hotter dust since the dust in such systems is near sites of active star formation or active nuclei.  Moreover, galaxies that appear as several bright clumps in the infrared such as Holmberg~II provide a large number of low optical depth lines-of-sight from which ultraviolet photons may escape (or their ultraviolet emission does not come from a single burst, but is rather more continuous or multi-generational in nature).  Such clumpy galaxies would hence show comparatively low infrared-to-ultraviolet ratios.  On the other hand, ultraviolet photons from galaxies that appear in the infrared as a single point-like blob of nuclear emission (e.g., NGC~1266) would encounter significant extinction, and hence such galaxies would exhibit high infrared-to-ultraviolet ratios and high dust temperatures.  In contrast to hot dust systems, galaxies with relatively low $f_\nu(70\micron)/f_\nu(160\micron)$ ratios have cooler dust because the dust is not in spatial proximity of the hot stars (e.g., Panagia 1973).  The heating of dust via the weaker ambient interstellar radiation field would be fractionally higher in these galaxies.  Therefore, their morphological appearance in the infrared should be comparatively smooth (e.g., NGC~2841).  

Since the relative distribution of interstellar grains and their heating sources is central to the scenario outlined above, we turn to the 24\m\ morphology of SINGS galaxies to provide a test of the above scenario.  MIPS 24\m\ data may be uniquely suited for such a test, as the data have significantly higher spatial resolution than either 70 or 160\m\ imaging (the 6\arcsec\ 24\m\ beam corresponds to $\sim$0.3~kpc at 10~Mpc), and effectively trace both interstellar grains and active sites of star formation\footnote{Note that the 24\m\ emission contains up to $\sim$25\% stellar emission for many SINGS early-type galaxies.} (e.g., Hinz et al. 2004; Gordon et al. 2004).  In fact, the 24\m\ emission can be spatially closely associated with \HII\ regions, and in such cases is probably dominated by dust from {\it within} these regions (Helou et al. 2004; Murphy et al. 2006).  To facilitate our analysis, we have decomposed the 24\m\ images into unresolved (point sources) and resolved emission.  The point source photometry 
is done using StarFinder (Diolaiti et al. 2000), which is appropriate for the stable and well sampled MIPS 24\m\ PSF.  A STinyTim (Krist 2002) model PSF with a temperature of 100~K, smoothed to account for pixel sampling, is used.  Smoothed STinyTim PSFs are excellent matches to observed MIPS 24\m\ PSFs (Engelbracht et al. 2006, in preparation).  
An image of all the detected point sources is created along with a difference image made by subtracting the point source image from the observed image.  The fluxes are measured in the point source (``unresolved'') and difference (``resolved'') images in the same aperture used for the total galaxy measurement (see Figure~\ref{fig:mosaic}).  In addition, nuclear fluxes are measured in a 12\arcsec\ radius circular aperture on the observed image.

\begin{figure}
 \plotone{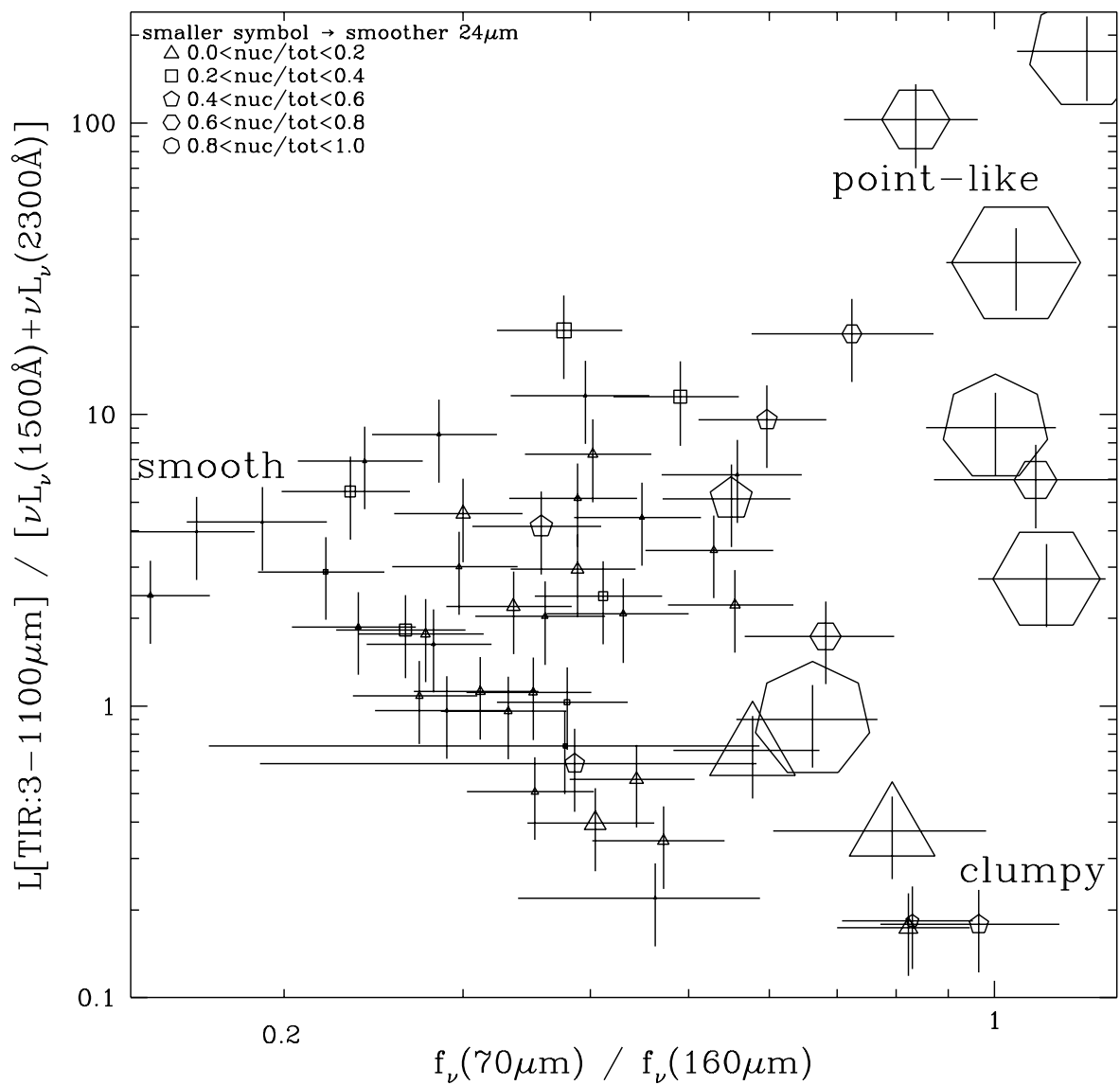}
 \caption{Similar to Figure~\ref{fig:ircolor}, but with symbol size scaled according to the ratio of unresolved-to-resolved  24\m\ emission; the largest symbols have this ratio equal to $\sim$10.  Each data point is also symbolized according to the ratio of nuclear-to-total 24\m\ emission (see Section~\ref{sec:geometry}).}
 \label{fig:ircolorb}
\end{figure}

\begin{figure}
 \plotone{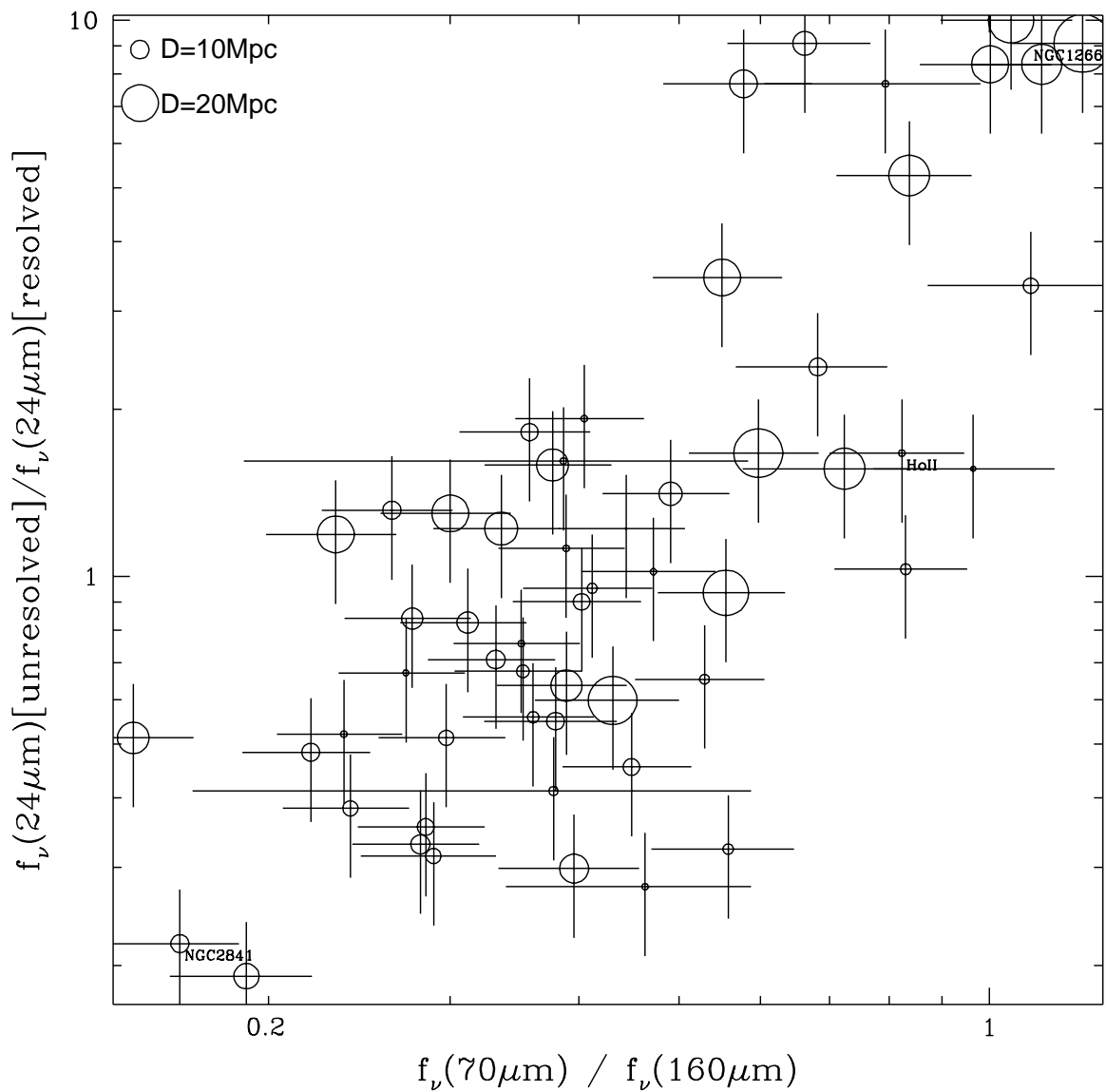}
 \caption{The ratio of unresolved-to-resolved 24\m\ emission as a function of far-infrared color (see Section~\ref{sec:geometry}).  A 25\% uncertainty is used for the error bars in the unresolved-to-resolved ratio.  The symbol sizes are scaled according to galaxy distance (see legend).}
 \label{fig:ircolorc}
\end{figure}

The results from this analysis are displayed in Figures~\ref{fig:ircolorb} and \ref{fig:ircolorc}.  In Figure~\ref{fig:ircolorb} the symbol size linearly scales with the ratio of unresolved-to-resolved 24\m\ emission, with the largest symbols corresponding to ratios $\sim$10.
In addition, each data symbol reflects the ratio of nuclear-to-total 24\m\ emission, as indicated in the figure legend.
Galaxies dominated by a single point source of nuclear emission at 24\m\ 
appear preferentially in the upper righthand portion of the diagram.  These galaxies contain hot dust and show relatively high infrared-to-ultraviolet ratios since the dust is centrally concentrated near the heating sources in the nuclei.  Note that nuclear activity is not the main factor in determining the 24\m\ morphology---only two of the point-like systems have active nuclei (NGC~1266 and NGC~5195).  Systems with clumpy 24\m\ morphologies appearing in the lower righthand corner 
still contain hot dust; the dust is concentrated around several heating sources, not just the nuclear ones.  Moreover, the clumpy distribution provides a larger number of low $\tau$ or `clean' lines-of-sight for ultraviolet photons to escape the galaxies, decreasing their infrared-to-ultraviolet ratios (see, for example, Roussel et al. 2005).  Finally, galaxies with smoother 24\m\ morphologies 
exhibit cooler far-infrared colors.  To see this latter effect more clearly, we show in Figure~\ref{fig:ircolorc} the ratio of unresolved-to-resolved 24\m\ emission as a function of far-infrared color.  Clearly there is a trend, indicating that the 24\m\ morphology can, for nearby galaxies, indicate the relative separation between interstellar grains and their heating sources.  
Note that distance contributes but does not dominate as a driver for the effect (the symbol sizes are scaled according to galaxy distance).

\subsection {Specific Star Formation Rate}
\label{sec:ssfr}

One way to parametrize the star formation history of a galaxy is via the star formation rate per stellar mass, or the specific star formation rate (SSFR).  Drory et al. (2004) and Feulner et al. (2005), for example, have utilized the specific star formation rate to explore the role of star formation in the growth of stellar mass over cosmic timescales.  In this work the specific star formation rate is quantified as a combination of the observed infrared and far-ultraviolet luminosities:
\be
{\rm SSFR~[yr^{-1}]} \simeq (~4.5{\rm TIR}[10^{37}{\rm W}] + 7.1\nu L_\nu(1500\AA)[10^{37}{\rm W}]~)~/~0.8~\nu L_\nu(K{\rm_s})[L_\odot]
\label{eqn:ssfr}
\ee
based on star formation rate conversion factors from Kennicutt (1998).  The numerator in Equation~\ref{eqn:ssfr}, applicable for galaxies with continuous star formation occurring on time scales $\gtrsim10^8$~yr, is a more robust way to quantify the star formation rate than relations that are limited to either infrared or ultraviolet luminosities.
The infrared luminosity accurately corresponds to the star formation rate only in the limiting case where all the star formation-related stellar emission is captured by interstellar dust grains.  Similarly, the ultraviolet emission can also be a poor measure of the star formation rate, especially when extinction is significant.  However, combining both the ultraviolet and infrared luminosities in Equation~\ref{eqn:ssfr} is akin to an extinction-corrected ultraviolet luminosity and thus more effectively recovers the true star formation rate (see also Bell 2003 and Iglesias-P\'aramo et al. 2006).  We note that similar values are obtained when using the optimal infrared$+$ultraviolet star formation rate tracer determined by Hirashita, Buat, \& Inoue (2003; their Equation~25), which incorporates parameters that account for the fraction of Lyman continuum luminosity ($f\simeq0.57$), the fraction of ultraviolet luminosity absorbed by dust ($\epsilon\simeq0.53$), and the fraction of dust heating by stellar populations older than $10^8$~yr ($\eta\simeq0.40$).  The $K{\rm_s}$ band luminosity in the denominator of Equation~\ref{eqn:ssfr} is equivalent to a stellar mass.  Bell et al. (2003), for example, fit stellar population synthesis models to thousands of 2MASS plus Sloan Digital Sky Survey optical-near-infrared datasets and find the distribution of $M_\ast/L_K{\rm_s}$ peaks near $\sim0.8~M_\odot/L_\odot$ for a wide range of galaxy masses.  Gavazzi, Pierini, \& Boselli (1996) also show that the dynamical mass in a galaxy is proportional to the $H$ band luminosity.

Figure~\ref{fig:ssfr} presents the interplay between the specific star formation rate, the infrared-to-ultraviolet ratio, and optical morphology.
With the exception of a handful of nuclear 24\m\ sources with high infrared-to-ultraviolet ratios, the SINGS sample shows a general trend in this diagram.  Galaxies with low specific star formation rates (SSFR$\lesssim$0.9~yr$^{-1}$) are of E, S0, S0/a, or Sa morphologies, consistent with the traditional notion that early-type galaxies exhibit low star formation rates per unit stellar mass (see, for example, Sandage 1986).  These early-type galaxies show increasing infrared-to-ultraviolet ratios for increasing specific star formation rates.  In contrast, spiral galaxies generally show SSFR$\gtrsim$0.9~yr$^{-1}$, and the later the spiral Hubble type, the larger the specific star formation rate and the smaller the infrared-to-ultraviolet ratio.  Note that the numerator in Equation~\ref{eqn:ssfr} overestimates the star formation rate for early-type galaxies, since the bulk of their infrared and ultraviolet luminosities are not due to recently-formed stars.


Assuming Equation~\ref{eqn:ssfr} can be reasonably applied to the SINGS early-type galaxies, in order to observe larger infrared-to-ultraviolet ratios (and thus larger dust extinction), increases in the specific star formation must be associated with increased amounts of interstellar dust.  On the other hand, increasing the specific star formation rate in late-type galaxies results in smaller infrared-to-ultraviolet ratios---the additional ultraviolet photons in spirals with high SSFRs tend to more easily escape the galaxies, since their clumpy distribution of dust provides many more sightlines of low optical depth than found in 24\m-smooth early-types.  In other words, the increased star formation rate in later-type spirals must lead to a higher density of holes through which ultraviolet photons can escape.  



\subsection {Luminosity}

\begin{figure}
 \plotone{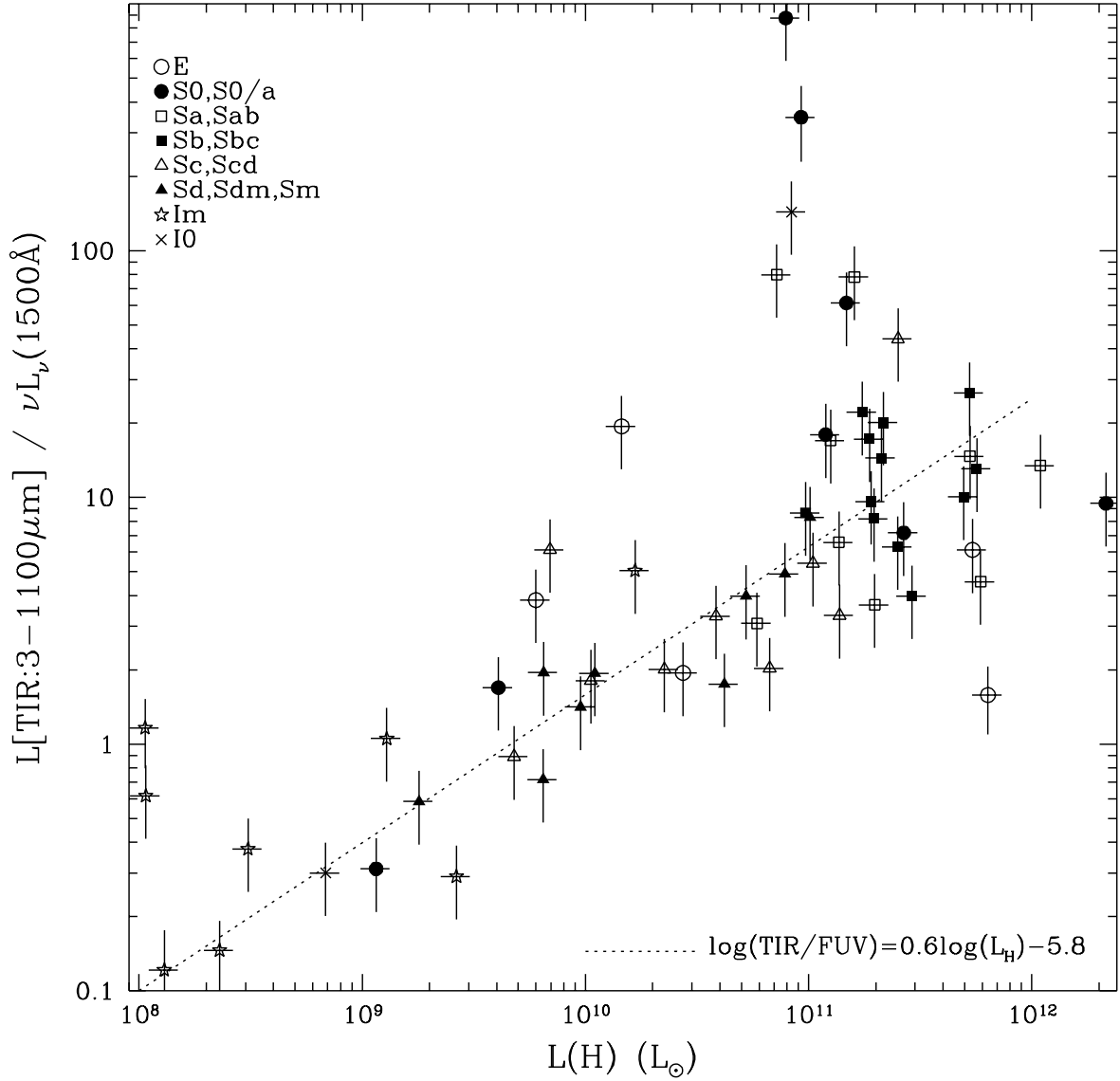}
 \caption{The infrared-to-far-ultraviolet ratio as a function of $H$ band luminosity.  The dotted line is a fit ``by eye'' to the general trend.}
 \label{fig:Hband}
\end{figure}

Global parameters related to galaxy structure, star formation history, molecular and atomic gas content, metallicity, etc. are known to trend with $H$ band luminosity, another popular proxy for galaxy stellar mass, especially for late-type galaxies (e.g., Gavazzi et al. 1996; Boselli et al. 2001).  Figure~\ref{fig:Hband} displays the infrared-to-far-ultraviolet ratio versus 2MASS $H$ band luminosity.  A clear correlation is found; more luminous (massive) galaxies show larger infrared-to-far-ultraviolet ratios, consistent with the findings of Cortese et al. (2006).  Though most of the low luminosity dwarfs and early-type galaxies follow the general trend, the data for E and S0/S0a galaxies contribute to increased scatter at high luminosity.  This increased scatter for early-type galaxies is not surprising given the large dispersion for these types of galaxies seen in Figure~\ref{fig:morphology}, and since the ultraviolet-emitting stars in the most massive early-types are generally associated with an old stellar population (see Boselli et al 2005).  Although the older stars might also contribute to the dust heating in massive early-type galaxies, the spectral shapes in Figures~\ref{fig:seds1}-\ref{fig:seds8} suggest that the bulk of the dust heating in these systems is dominated by intermediate-age stars emitting mostly in the visible.  In contrast, the spectral shapes for most later-type galaxies in Figures~\ref{fig:seds1}-\ref{fig:seds8} indicate that the bulk of the dust heating is being carried out by a younger (bluer) stellar population.

\subsection {Ultraviolet Spectral Slope}
\label{sec:beta}

The infrared-to-ultraviolet ratio has been shown to be fairly tightly correlated with the ultraviolet spectral slope in starburst galaxies, an important discovery that allows the extinction at ultraviolet wavelengths to be estimated from ultraviolet spectral data alone (e.g., Calzetti, Kinney, \& Storchi-Bergmann 1994; Calzetti 1997; Meurer, Heckman, \& Calzetti 1999).  A starburst galaxy is defined here as a galaxy experiencing prodigious, recent star formation (perhaps triggered by an encounter) at a rate that cannot be sustained over the lifetime of the galaxy.  Non-starbursting galaxies have also been studied in the context of infrared-to-ultraviolet ratio and ultraviolet spectral slope, but their data show a larger dispersion, with normal star-forming and quiescent systems exhibiting redder ultraviolet spectra and/or lower infrared-to-ultraviolet ratios (e.g., Buat et al. 2002; Bell 2002; Kong et al. 2004; Gordon et al. 2004; Buat et al. 2005; Burgarella, Buat, \& Iglesias-P\'aramo 2005; Calzetti et al. 2005; Seibert et al. 2005; Cortese et al. 2006; Boissier et al. 2006; Gil de Paz et al. 2006).  The intrinsic ultraviolet spectral slope is quite sensitive to the effective age of the stellar population, leading Calzetti et al. (2005) to suggest that the evolved, non-ionizing stellar population ($\sim$50-100~Myr) dominates the ultraviolet emission in normal systems, in contrast to current star formation processes dominating the ultraviolet emission in starbursts.  The increased diversity in the ultraviolet spectral slopes for evolved stellar populations manifests itself as a larger dispersion for quiescent and normal star-forming galaxies in plots of the infrared-to-ultraviolet ratio as a function of ultraviolet spectral slope.  Interestingly, Boissier et al. (2006) use azimuthally-averaged radial profiles, and after excluding emission from the bulge/nucleus, they find the relation between infrared-to-ultraviolet and ultraviolet slope tightens up compared with the one obtained using the integrated data.  This result is consistent with the interpretation of Calzetti et al. if either the evolved stellar populations in normal star-forming galaxy bulges cause the increased scatter compared to the starburst trend, or if azimuthally averaging smooths over small scale effects such as the heating of dust in one region by ultraviolet light from a different nearby region.

Figure~\ref{fig:beta} displays a diagram of spatially-integrated infrared-to-ultraviolet ratios and ultraviolet spectral slopes.  Normal star-forming and starbursting galaxies from Kong et al. (2004) and Calzetti et al. (1995) are plotted in addition to the SINGS data points.  The dotted curve is that for starbursting galaxies from Kong et al. (2004) and the solid curve is applicable to normal star-forming galaxies of type Sa or later (Cortese et al. 2006).  Similar to what has been found for other samples of non-starbursting galaxies, the SINGS dataset shows more scatter in this diagram and the galaxies are redder in their ultraviolet spectral slope compared to starburst galaxies.  Inspection of the distribution as a function of SINGS optical morphology, however, shows that the 14 reddest SINGS galaxies are type Sab or earlier, a result that is perhaps expected for systems with quenched star formation rate histories; the early-type galaxies in SINGS contribute to most of the observed scatter, such that the trend for the subset of just late-type SINGS galaxies shows a dispersion comparable to that for starbursts.  The 14 reddest SINGS galaxies also significantly differ from even the normal galaxy curve, but this is consistent with the fact that Cortese et al. excluded types earlier than Sa for their analysis.  Finally, though the SINGS sample shows very large dispersion and does not as a whole match the starburst trend, we have verified the starburst subset of SINGS (e.g., Markarian~33, NGC~1705, NGC~2798, NGC~3034, NGC~7552) does indeed match the canonical starburst trend.


\section {Discussion and Summary}
\label{sec:summary}

The ultraviolet-to-radio broadband spectral energy distributions are presented for the 75 galaxies in the \Spitzer\ Infrared Nearby Galaxies Survey, a collection of galaxies that broadly samples the wide variety of galaxy morphologies, luminosities, colors, and metallicities seen in the Local Universe.  The infrared-to-ultraviolet ratio is explored in conjunction with several global parameters.  An interesting empirical finding is that systems with cooler dust show a restricted range of infrared-to-ultraviolet ratios ($\sim$0.5~dex), while systems with warm global far-infrared colors exhibit a large range of infrared-to-ultraviolet ratios ($\sim$3~dex).  To put it another way, the cold dust sytems in the SINGS sample show average ultraviolet extinctions; no cold galaxy is particularly optically thick or thin.  There remains the possibility that part of this distribution is attributable to selection effects, but we use the morphology from MIPS 24\m\ imaging to interpret this distribution to result from the relative distribution of dust grains and their heating sources.  Nearby galaxies with globally cooler dust appear smoother at 24\m, from which we infer that the dust grains are well mixed throughout the interstellar medium and not concentrated near sites of active star formation.  On the other hand, galaxies with elevated \Spitzercolora\ ratios appear as one or a handful of clumps at 24\m\ and thus have much of their dust considerably closer to heating sources.  The observed range in infrared-to-ultraviolet ratio is also related to the 24\m\ morphology, from which the density of available clean lines-of-sight for ultraviolet photons to escape can be inferred.  The dust distribution in galaxies appearing as a single clump at 24\m\ heavily enshrouds the heating sources (high infrared-to-ultraviolet ratios), galaxies with multiple clumps at 24\m\ provide a large number of low optical depth lines-of-sight along which ultraviolet photons can escape (low infrared-to-ultraviolet ratios), and a smooth distribution at 24\m\ implies a dust distribution that provides an intermediate number of low optical depth lines-of-sight (average infrared-to-ultraviolet ratios).  Detailed studies of the relative distributions of the infrared emission and the ionizing radiation fields in SINGS galaxies have been carried out in IC~2574 (Cannon et al. 2005), NGC~1705 (Cannon et al. 2006a), and NGC~6822 (Cannon et al. 2006b).  These dwarf galaxies appear as multiple clumps at 24\m\ and show low optical extinctions and highly variable ratios of H$\alpha$-to-infrared (i.e., significant ultraviolet photon leakage), consistent with our expectation that multi-clump 24\m\ galaxies should have warm far-infrared colors and low global infrared-to-ultraviolet ratios.

\begin{figure}
 \plotone{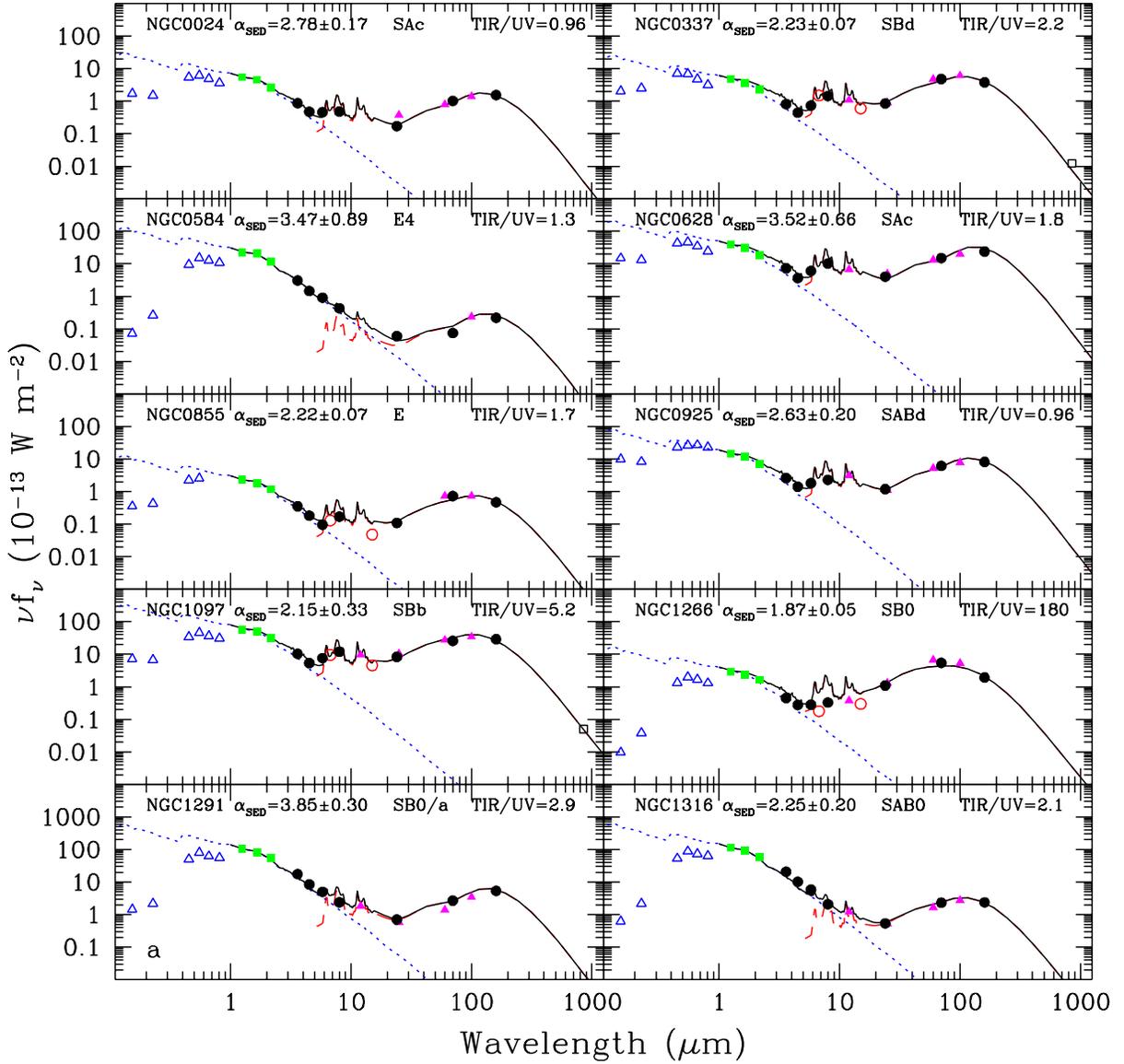}
 \caption{Globally-integrated 0.15-850\m\ spectral energy distributions for the SINGS sample.  \GALEX\ and optical, 2MASS, \Spitzer, \IRAS, \ISO, and SCUBA data are represented by open triangles, filled squares, filled circles, filled triangles, open circles, and open squares, respectively.  The solid curve is the sum of a dust (dashed) and a stellar (dotted) model.  The dust curve is a Dale \& Helou (2002) model fitted to ratios of the 24, 70, and 160\m\ fluxes; the $\alpha_{\rm SED}$ listed within each panel parametrizes the distribution of dust mass as a function of heating intensity, as described in 
Dale \& Helou (2002).  The stellar curve is a 1~Gyr continuous star formation, solar metallicity curve from Vazquez \& Leitherer (2005) fitted to the 2MASS data (see \S~\ref{sec:seds} for details).}
 \label{fig:seds1}
\end{figure}

\begin{figure}
 \plotone{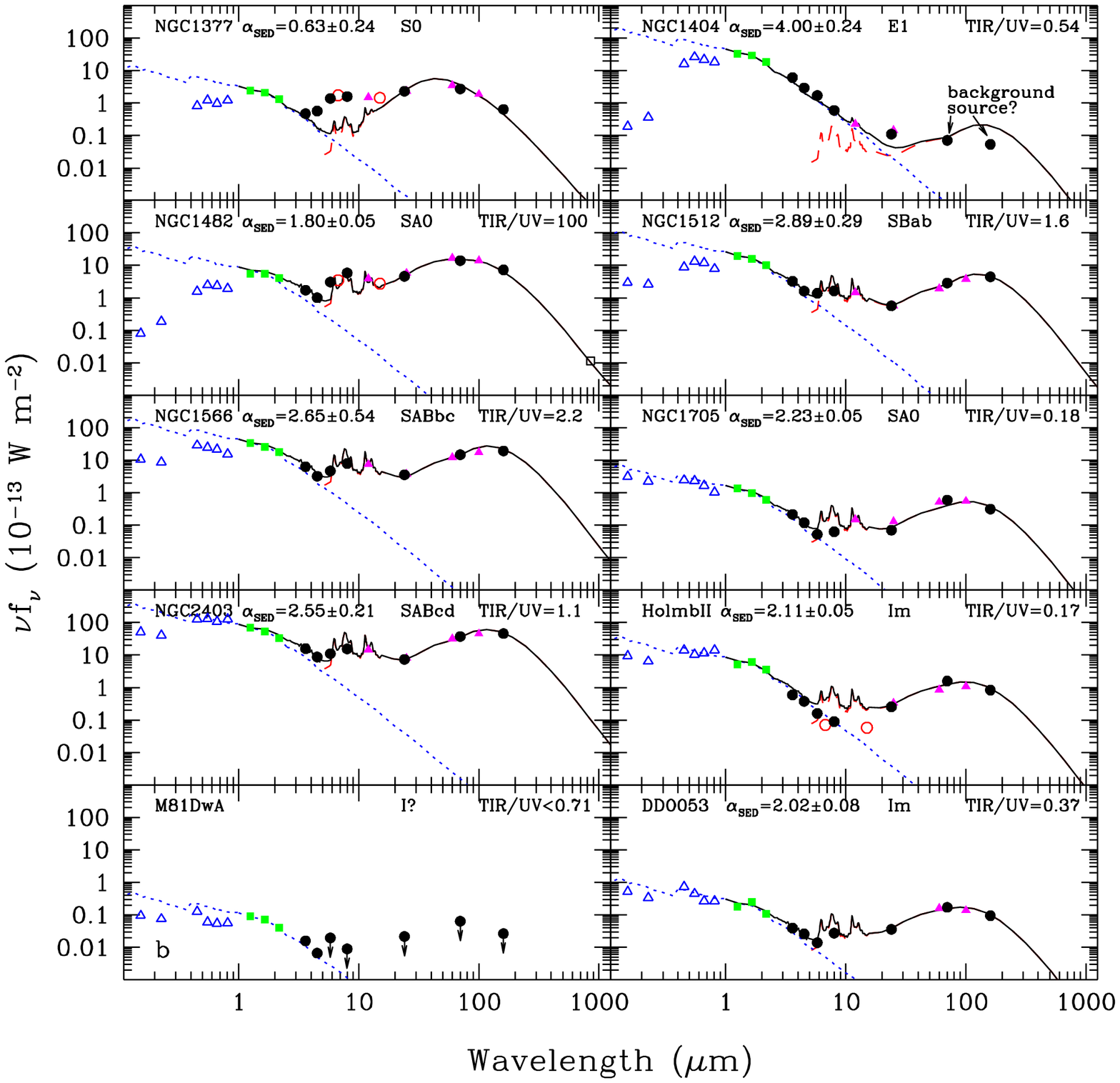}
 \caption{Globally-integrated 0.15-850\m\ spectral energy distributions for the SINGS sample (continued).}
 \label{fig:seds2}
\end{figure}

\begin{figure}
 \plotone{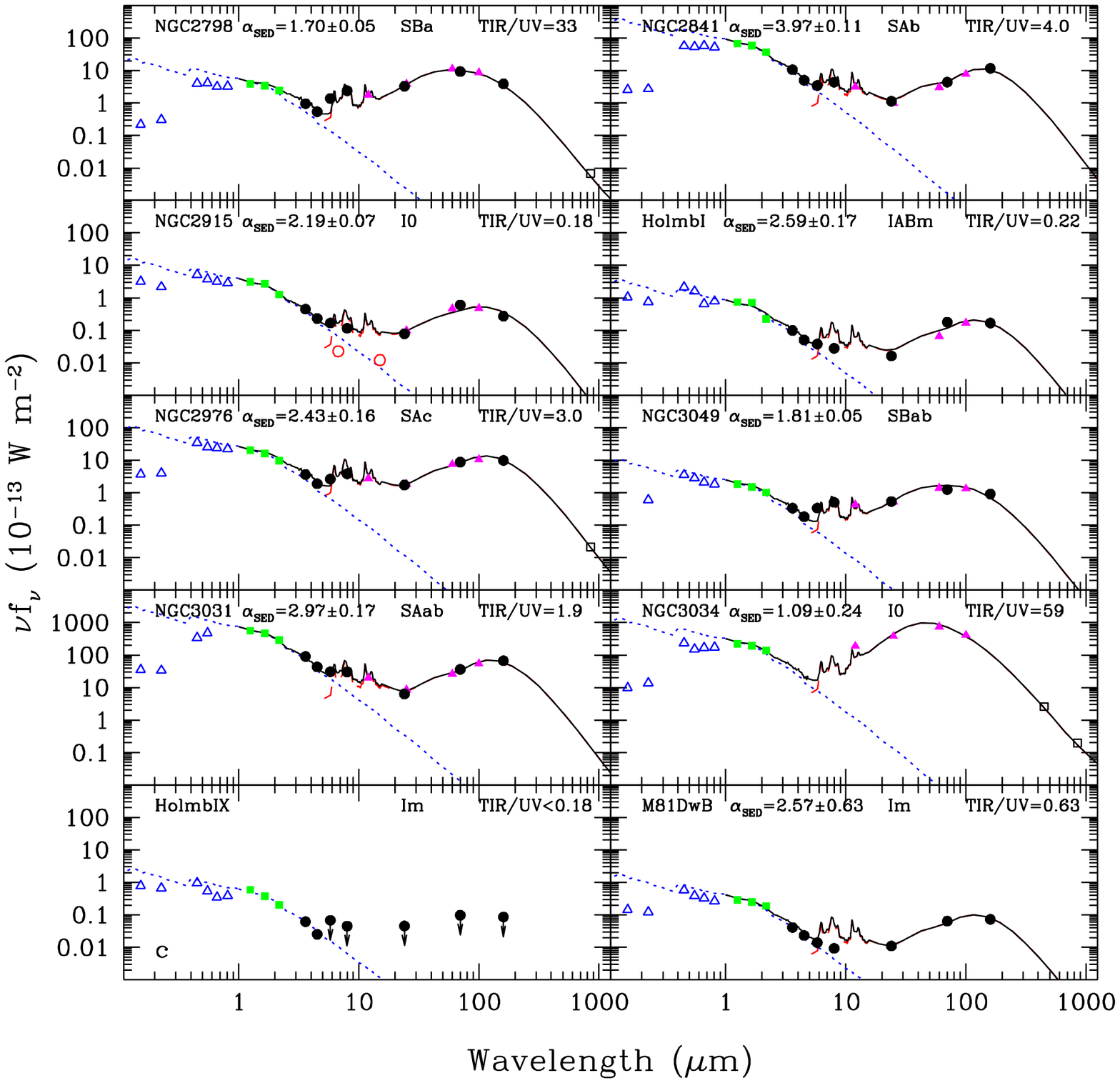}
 \caption{Globally-integrated 0.15-850\m\ spectral energy distributions for the SINGS sample (continued).}
 \label{fig:seds3}
\end{figure}

\begin{figure}
 \plotone{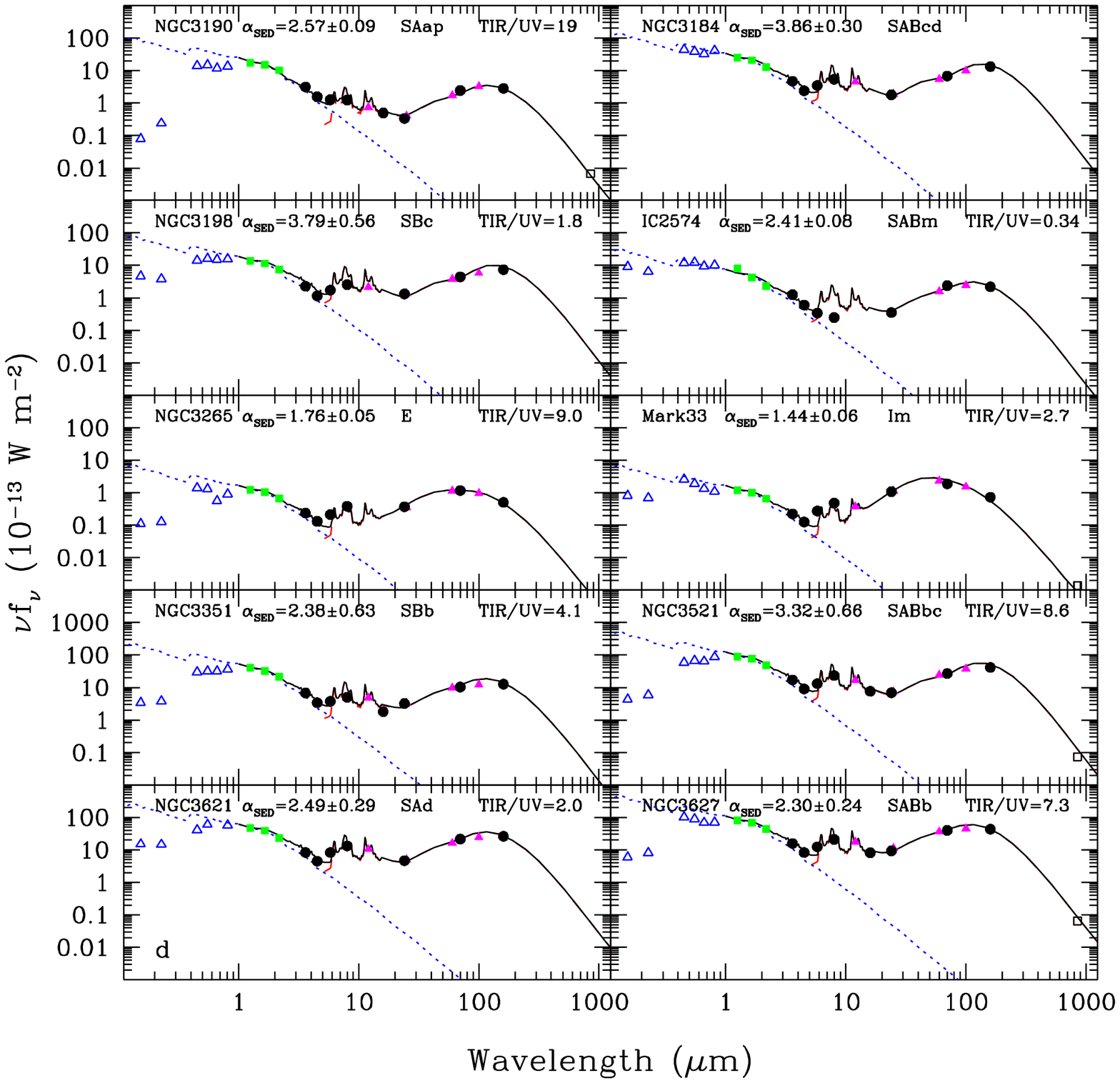}
 \caption{Globally-integrated 0.15-850\m\ spectral energy distributions for the SINGS sample (continued).}
 \label{fig:seds4}
\end{figure}

\begin{figure}
 \plotone{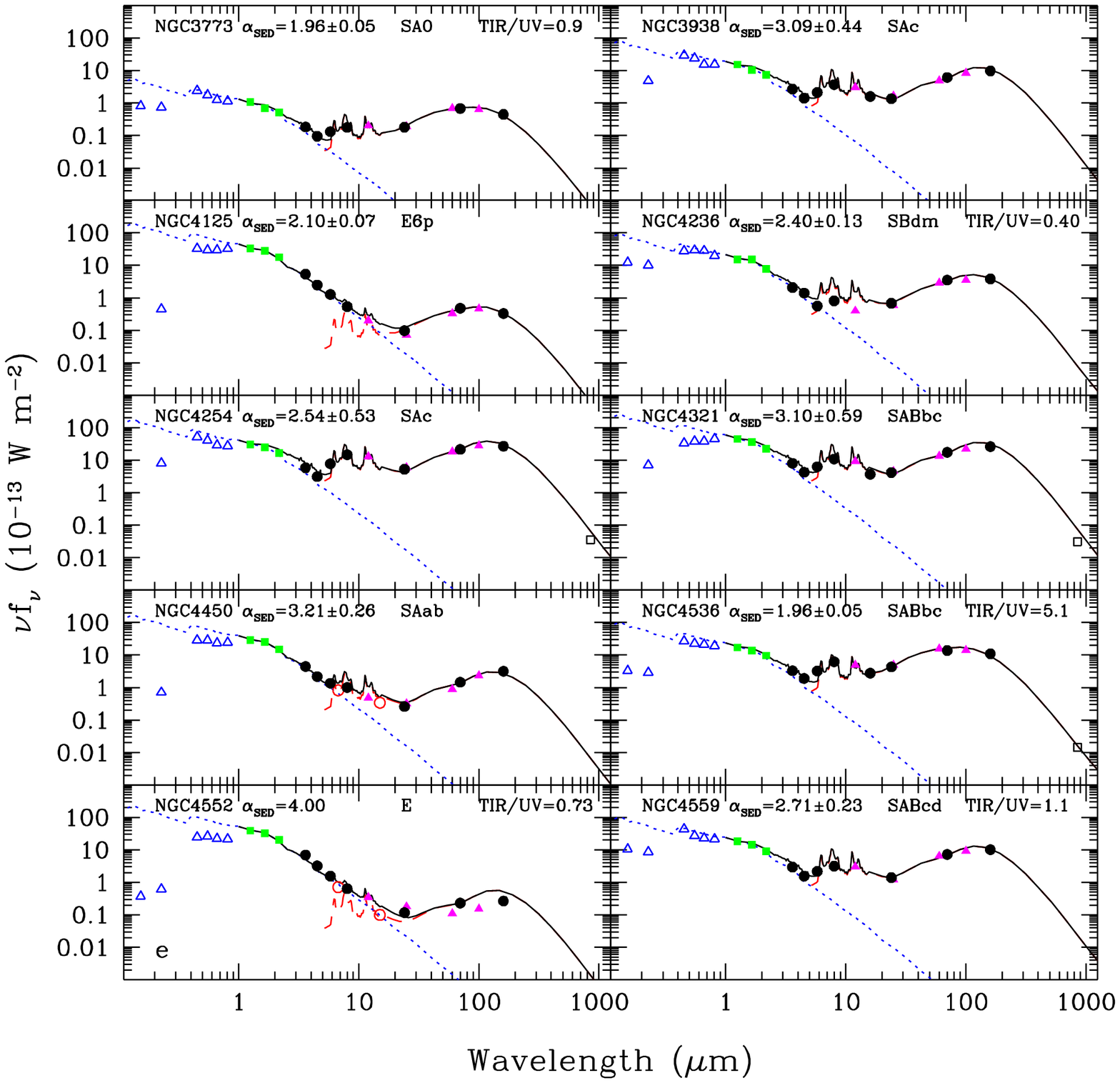}
 \caption{Globally-integrated 0.15-850\m\ spectral energy distributions for the SINGS sample (continued).}
 \label{fig:seds5}
\end{figure}

\begin{figure}
 \plotone{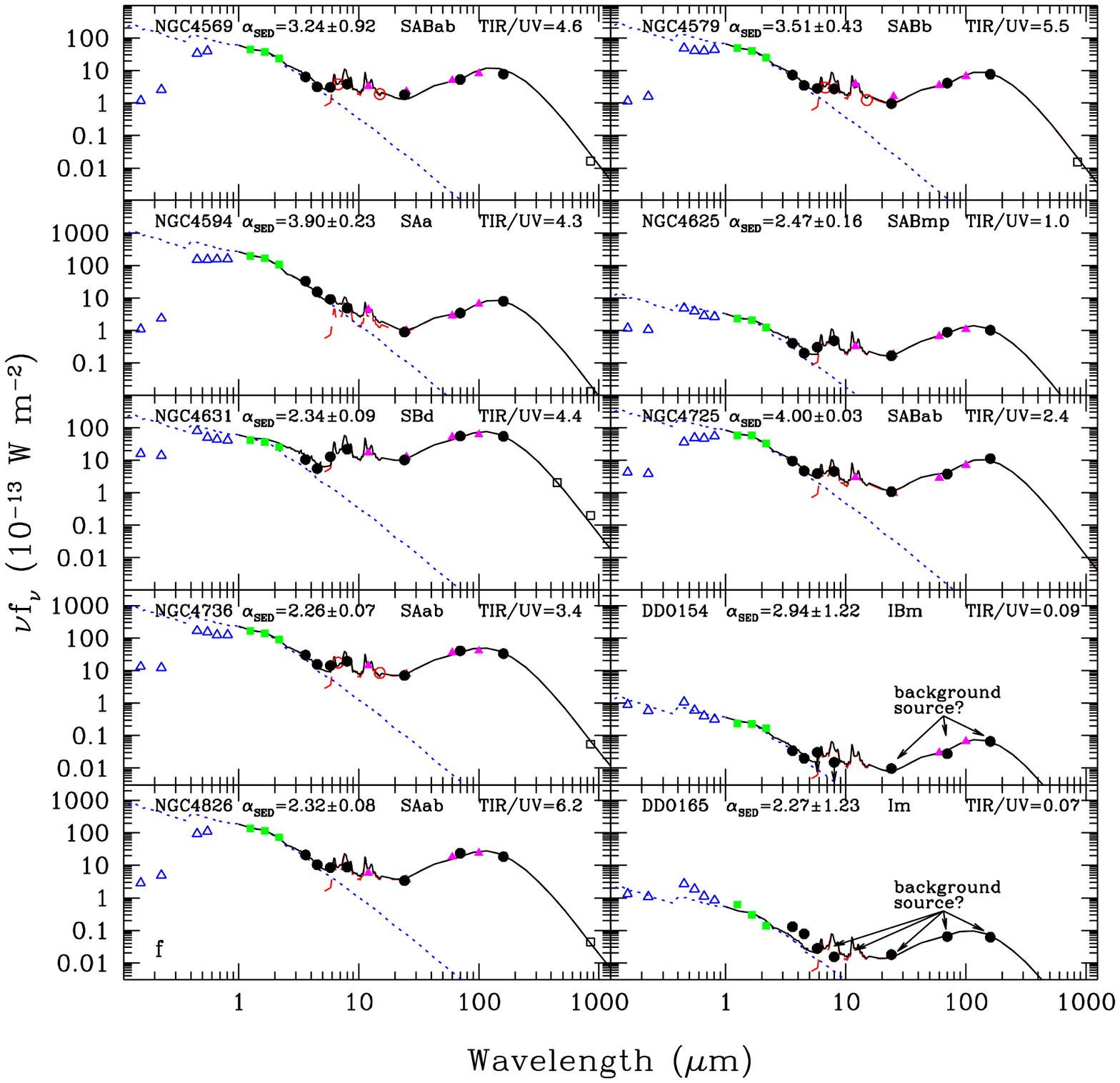}
 \caption{Globally-integrated 0.15-850\m\ spectral energy distributions for the SINGS sample (continued).}
 \label{fig:seds6}
\end{figure}

\begin{figure}
 \plotone{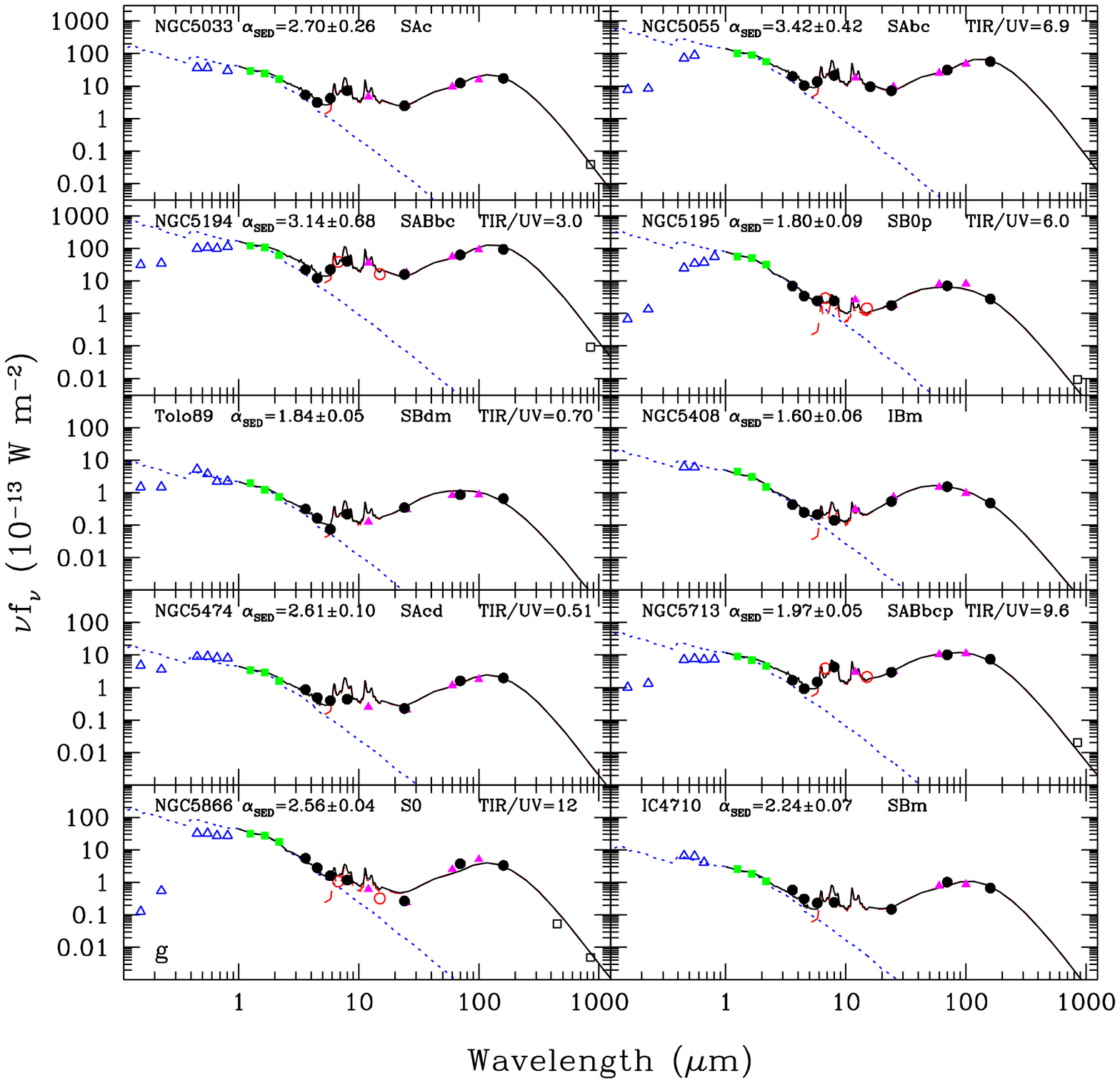}
 \caption{Globally-integrated 0.15-850\m\ spectral energy distributions for the SINGS sample (continued).}
 \label{fig:seds7}
\end{figure}

\begin{figure}
 \plotone{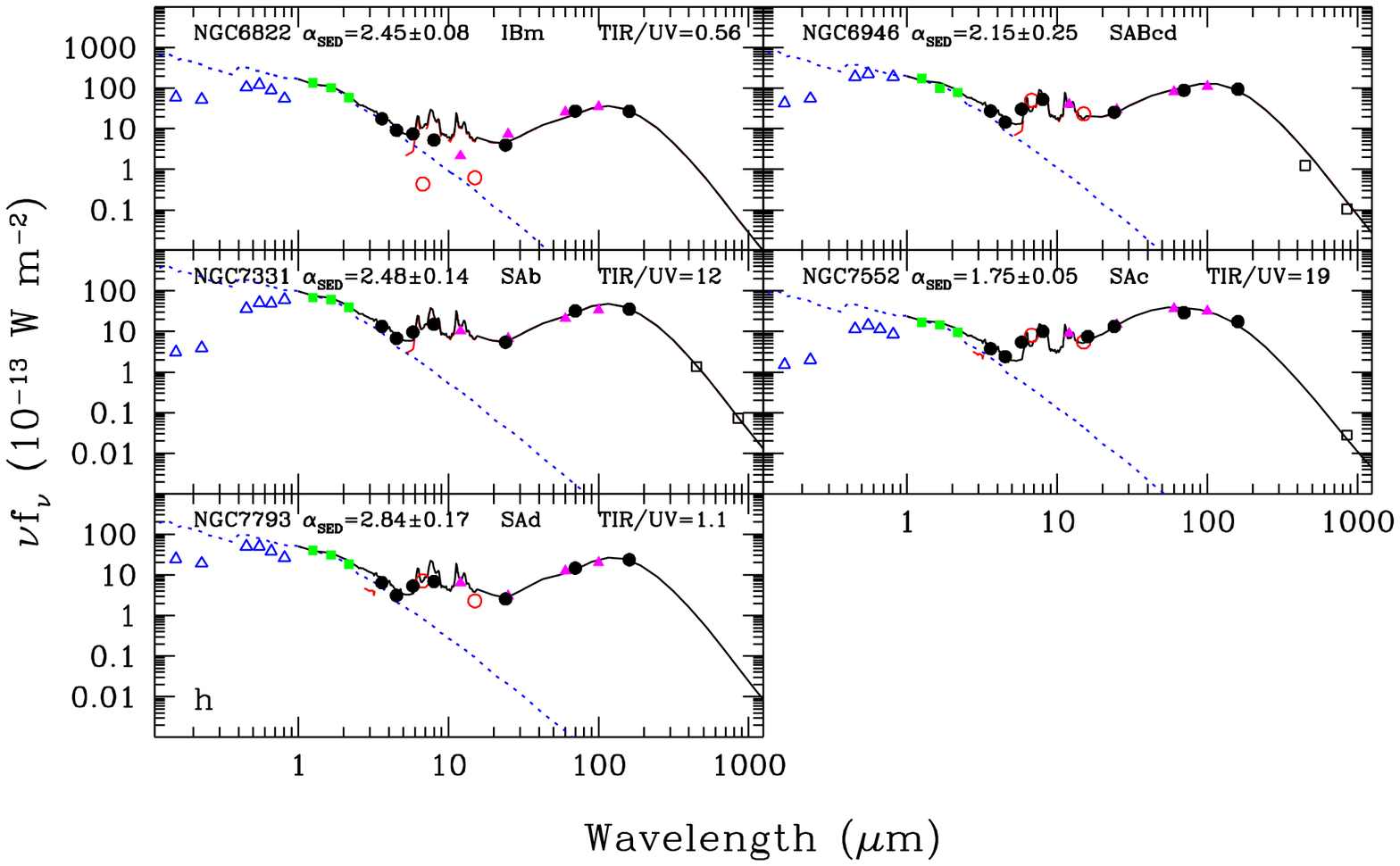}
 \caption{Globally-integrated 0.15-850\m\ spectral energy distributions for the SINGS sample (continued).}
 \label{fig:seds8}
\end{figure}

\begin{figure}
 \plotone{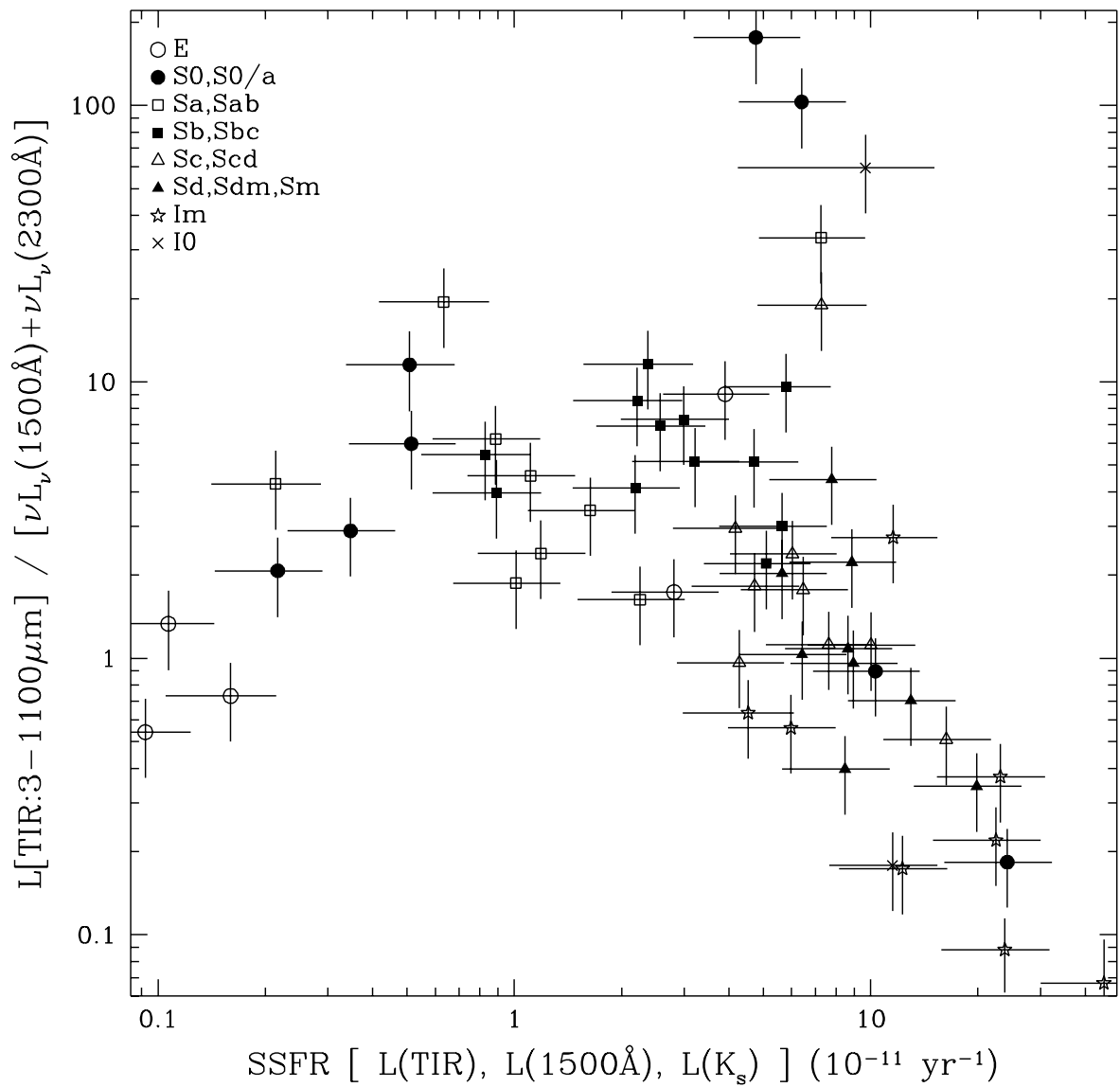}
 \caption{The infrared-to-ultraviolet ratio as a function of the specific star formation rate (Equation~\ref{eqn:ssfr}).  The error bars derive from the observational uncertainties plus a 30\% factor assumed for converting the $K_{\rm s}$ luminosity to a stellar mass (see Section~\ref{sec:ssfr}).}
 \label{fig:ssfr}
\end{figure}

\begin{figure}
 \plotone{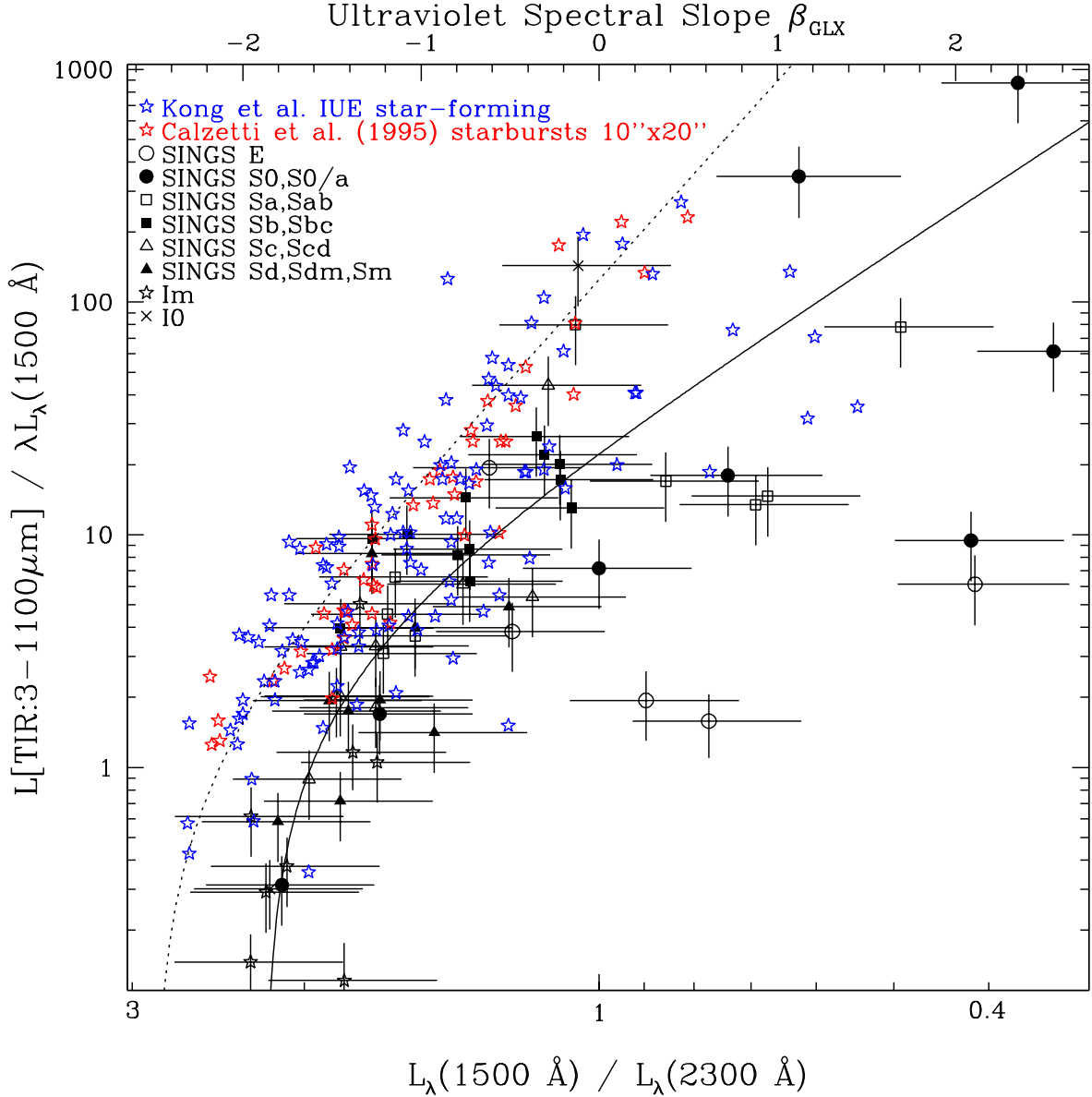}
 \caption{The infrared-to-far-ultraviolet ratio as a function of ultraviolet spectral slope.  Normal star-forming and starbursting galaxies from Kong et al. (2004) and Calzetti et al. (1995) are plotted in addition to the SINGS data points.  The dotted curve is that for starbursting galaxies from Kong et al. (2004) and the solid curve is applicable to normal star-forming galaxies (Cortese et al. 2006).}
 \label{fig:beta}
\end{figure}

A principal component analysis of the SINGS broadband spectra indicates that most of the sample's large broadband spectral variations stem from two underlying components, one typical of a galaxy with a low infrared-to-ultraviolet ratio (88\% of the sample variation) and one indicative of a galaxy with a high infrared-to-ultraviolet ratio (7\% of the sample variation).  The implication is that the star formation history (i.e., the specific star formation rate, the birthrate parameter or some other measure of the current-to-past star formation rate) may be the dominant regulator of the broadband spectral variations between galaxies.  From a morphological standpoint, we find that much of the dispersion in plots such as infrared-to-ultraviolet versus ultraviolet spectral slope (Figure~\ref{fig:beta}) stems from early-type galaxies, which have significantly redder ultraviolet spectra than other galaxy types.  In fact, the galaxies with the highest optical-to-infrared ratios, the smallest specific star formation rates, and the reddest ultraviolet slopes are all early-type galaxies (see Figures~\ref{fig:seds1}-\ref{fig:seds8}, \ref{fig:ssfr}, and \ref{fig:beta}, respectively).  

 Evidence for the star formation history regulating spectral variations is found in a striking trend in the infrared-to-ultraviolet ratio as a function of the specific star formation rate (Figure~\ref{fig:ssfr}).  Early-type galaxies show higher ratios of infrared-to-ultraviolet (higher dust extinction) for larger specific star formation rates, implying that the specific star formation rate in ellipticals and S0 galaxies is closely tied to the {\it amount} of dust.  Conversely, spiral galaxies show lower infrared-to-ultraviolet ratios (lower dust extinction) for higher specific star formation rates, suggesting that the specific star formation rate in ellipticals is linked to the {\it distribution} of dust---in spiral galaxies a larger number of holes are created for increased star formation activity, holes through which ultraviolet light more easily passes out of galaxies.

In a study of 99,088 galaxies from the Sloan Digital Sky Survey, Obri\'c et al. (2006) find that the GALEX, Sloan, and 2MASS data ``form a nearly one parameter family.'' In particular, they can predict with 20\% accuracy the 2MASS $K_{\rm s}$ flux using just the Sloan $u$ and $r$ fluxes.  In addition, they can predict to within a factor of two certainty the $IRAS$ 60\m\ flux based on the Sloan broadband data.  Such simple optical-infrared correlations are not seen for SINGS galaxies.  However, Obri\'c et al. are only able to identify $IRAS$ fluxes for less than 2\% of their sample, and this subset is strongly biased to optically blue galaxies.  The SINGS sample, though far smaller in size, provides complete panchromatic information for a far more diverse ensemble of galaxies and is thus much less biased to a particular subset of the local galaxy population.


\acknowledgements 
Support for this work, part of the {\it Spitzer Space Telescope} Legacy Science Program, was provided by NASA through Contract Number 1224769 issued by the Jet Propulsion Laboratory, California Institute of Technology under NASA contract 1407.  AGdP is financed by the MAGPOP EU Marie Curie Research Training Network and the Spanish Programa Nacional de Astronom\'{\i}a y Astrof\'{\i}sica under grant AYA2003-01676.  We are thankful for the hard work put in by the instrument teams and the \Spitzer\ Science Center.  We gratefully acknowledge NASA's support for construction, operation, and science analysis for the GALEX mission, developed in cooperation with the Centre National d'Etudes Spatiales of France and the Korean Ministry of Science and Technology.  This research has made use of the NASA/IPAC Extragalactic Database which is operated by JPL/Caltech, under contract with NASA.  This publication makes use of data products from the Two Micron All Sky Survey, which is a joint project of the University of Massachusetts and IPAC/Caltech, funded by NASA and the National Science Foundation.

\begin{deluxetable}{llcrrrccc}
\def\a{\tablenotemark{a}}
\def\b{\tablenotemark{b}}
\def\c{\tablenotemark{c}}
\def\d{$^\dagger$}
\def\p{$\pm$}
\tabletypesize{\scriptsize}
\tablenum{1}
\tablecaption{Galaxy Data \label{tab:additional_data}}
\tablewidth{0pc}
\tablehead{
\colhead{Galaxy} &
\colhead{Optical} &
\colhead{$\alpha_0$~\&~$\delta_0$} &
\colhead{2$a$} &
\colhead{2$b$} &
\colhead{PA} &
\colhead{$f_\nu(24)[{\rm unres}]/$} &
\colhead{$f_\nu(24)[{\rm nuc}]/$} &
\colhead{$\theta_{\rm FUV}$\b}
\\
\colhead{} &
\colhead{Morph.} &
\colhead{(J2000)} &
\colhead{($\arcsec$)} &
\colhead{($\arcsec$)} &
\colhead{($\degr$)} &
\colhead{$f_\nu(24)[{\rm res}]$\a} &
\colhead{$f_\nu(24)[{\rm total}]$\a} &
\colhead{($\arcsec$)}
}
\startdata
NGC~0024\d     &SAc  &000955.9$-$245755& 301& 216&135&0.31    &0.14   &~36.0  \\
NGC~0337\d     &SBd  &005950.7$-$073444& 253& 194& 50&0.93    &0.17   &~26.7  \\
NGC~0584\d     &E4   &013120.6$-$065205& 326& 278&330&0.49    &0.39   &~28.8  \\
NGC~0628\d     &SAc  &013641.8$+$154717& 721& 717&248&0.84    &0.01   &150.~  \\
NGC~0855       &E    &021403.9$+$275239& 190& 170&338&2.38    &0.69   &\nodata\\
NGC~0925\d     &SABd &022713.6$+$333504& 735& 486& 15&0.71    &0.02   &114~~  \\
NGC~1097\d     &SBb  &024618.0$-$301642& 758& 612& 40&0.64    &0.12   &129~~  \\
NGC~1266\d     &SB0  &031600.7$-$022541& 234& 232&  0&9.09    &0.87   &~10.5  \\
NGC~1291\d     &SB0/a&031719.1$-$410632& 840& 803&  0&0.48    &0.21   &253~~  \\
NGC~1316\d     &SAB0 &032241.2$-$371210& 864& 583&230&0.60    &0.05   &~89.2  \\
NGC~1377       &S0   &033639.0$-$205408& 181& 162&  0&$\sim$20&0.85   &\nodata\\
NGC~1404       &E1   &033852.3$-$353540& 524& 369&239&0.57    &0.29   &\nodata\\
NGC~1482\d     &SA0  &035439.0$-$203009& 349& 310& 29&5.26    &0.77   &~19.0  \\
NGC~1512\d     &SBab &040355.0$-$432044& 491& 287&325&0.33    &0.10   &136~~  \\
NGC~1566\d     &SABbc&042000.4$-$545615& 552& 435& 40&1.22    &0.11   &~84.3  \\
NGC~1705\d     &SA0  &045413.5$-$532137& 167& 120&130&1.03    &0.43   &\nodata\\
NGC~2403\d     &SABcd&073655.0$+$653554&1164& 848& 40&0.76    &0.01   &161~~  \\
Holmberg~II\d  &Im   &081906.8$+$704309& 441& 430&  0&1.67    &0.01   &119~~  \\
M81~Dwarf~A    &I?   &082356.0$+$710145&  78&  78&  0&\nodata &\nodata&\nodata\\
DDO~053\d      &Im   &083406.8$+$661036& 133& 110& 30&7.69    &0.08   &~22.9  \\
NGC~2798\d     &SBa  &091723.1$+$415957& 235& 232&  0&$>$10   &0.75   &~~7.8  \\
NGC~2841\d     &SAb  &092203.3$+$505837& 550& 342&150&0.22    &0.04   &~74.2  \\
NGC~2915\d     &I0   &092609.4$-$763736& 183& 132&290&1.56    &0.53   &\nodata\\
Holmberg~I\d   &IABm &094030.5$+$711033& 265& 228&120&0.28    &0.01   &~59.9  \\
NGC~2976\d     &SAc  &094715.3$+$675507& 457& 311&322&1.12    &0.05   &~45.2  \\
NGC~3049       &SBab &095449.6$+$091614& 218& 160&119&5.88    &0.74   &\nodata\\
NGC~3031       &SAab &095531.8$+$690403&1628&1122&154&0.52    &0.07   &324~~  \\
NGC~3034\c     &I0   &\nodata          &\nodata&\nodata&\nodata&\nodata&\nodata&\nodata\\
Holmberg~IX    &Im   &095729.2$+$690250& 247& 180&130&\nodata &\nodata&~47.8  \\
M81~Dwarf~B\d  &Im   &100531.3$+$702152& 107&  69&140&1.61    &0.50   &\nodata\\
NGC~3190\d     &SAap &101805.7$+$214957& 334& 196&117&1.59    &0.35   &~29.2  \\
NGC~3184       &SABcd&101815.6$+$412542& 614& 538&349&0.73    &0.01   &\nodata\\
NGC~3198\d     &SBc  &101954.8$+$453301& 518& 315&125&1.32    &0.34   &\nodata\\
IC~2574\d      &SABm &102822.7$+$682448& 827& 376&140&1.02    &0.03   &188~~  \\
NGC~3265\d     &E    &103106.8$+$284751& 184& 175&320&8.33    &0.82   &\nodata\\
Markarian~33\d &Im   &103231.2$+$542359& 181& 177&  0&8.33    &0.75   &\nodata\\
NGC~3351\d     &SBb  &104357.5$+$114219& 586& 457& 10&1.82    &0.46   &~84.5  \\
NGC~3521\d     &SABbc&110548.7$-$000222& 766& 494&342&0.35    &0.04   &~89.6  \\
NGC~3621      &SAd   &111818.3$-$324855& 791& 555&340&0.56    &0.02   &125~~  \\
NGC~3627\d    &SABb  &112013.4$+$125927& 745& 486&347&0.90    &0.01   &~56.0  \\
NGC~3773\d    &SA0   &113813.1$+$120644&  96&  94&  0&9.09    &0.85   &\nodata\\
NGC~3938      &SAc   &115250.3$+$440715& 504& 468&  0&0.58    &0.04   &\nodata\\
NGC~4125      &E6p   &120805.8$+$651024& 228& 151&  0&0.56    &0.40   &\nodata\\
NGC~4236\d    &SBdm  &121635.9$+$692808&1129& 420&155&1.92    &0.004  &169~~  \\
NGC~4254      &SAc   &121849.7$+$142519& 519& 420&330&0.48    &0.03   &\nodata\\
NGC~4321      &SABbc &122254.8$+$154907& 558& 483&310&0.29    &0.09   &\nodata\\
NGC~4450      &SAab  &122830.1$+$170454& 401& 284&  0&0.57    &0.08   &\nodata\\
NGC~4536\d    &SABbc &123427.5$+$021113& 454& 376& 30&3.45    &0.48   &~79.4  \\
NGC~4552\d    &E     &123539.8$+$123323& 306& 306&  0&0.41    &0.52   &~13.3  \\
NGC~4559\d    &SABcd &123558.1$+$275752& 576& 327& 50&0.83    &0.04   &~87.1  \\
NGC~4569      &SABab &123650.2$+$131001& 593& 327& 21&1.30    &0.10   &~59.2  \\
NGC~4579\d    &SABb  &123743.6$+$114900& 295& 229&  0&1.19    &0.27   &~54.5  \\
NGC~4594\d    &SAa   &123959.4$-$113714& 554& 232&  0&0.19    &0.15   &~61.0  \\
NGC~4625\d    &SABmp &124152.3$+$411618& 198& 190&140&0.55    &0.21   &~28.1  \\
NGC~4631\d    &SBd   &124203.7$+$323205& 952& 539&350&0.45    &0.03   &~84.3  \\
NGC~4725\d    &SABab &125027.7$+$252948& 689& 523& 30&0.51    &0.01   &124~~  \\
NGC~4736\d    &SAab  &125056.7$+$410706&1033& 824& 10&0.65    &0.07   &~40.7  \\
DDO~154\d     &IBm   &125405.2$+$270854& 198& 126&123&1.16    &0.00   &~33.2  \\
NGC~4826      &SAab  &125642.8$+$214050& 722& 448&112&0.32    &0.14   &~41.2  \\
DDO~165\d     &Im    &130625.0$+$674226& 267& 150&  0&$>$10   &0.05   &~41.7  \\
NGC~5033      &SAc   &131328.2$+$363534& 729& 467&  0&0.36    &0.12   &\nodata\\
NGC~5055      &SAbc  &131548.3$+$420142& 893& 682& 11&0.38    &0.04   &108~~  \\
NGC~5194\d    &SABbc &132950.6$+$471307&1699&1129&285&0.51    &0.002  &143~~  \\
NGC~5195\d    &SB0p  &132959.4$+$471556& 202& 191&  0&3.32    &0.61   &110~~  \\
Tololo~89     &SBdm  &140121.3$-$330401& 196& 130&  0&7.69    &0.09   &~31.9  \\
NGC~5408      &IBm   &140321.1$-$412241& 256& 209& 67&3.57    &0.02   &\nodata\\
NGC~5474\d    &SAcd  &140459.9$+$533913& 386& 335&120&0.68    &0.04   &~84.3  \\
NGC~5713\d    &SABbcp&144011.2$-$001726& 153& 140&  0&1.67    &0.49   &~15.0  \\
NGC~5866\d    &S0    &150628.8$+$554551& 500& 306& 39&1.41    &0.34   &~26.3  \\
IC~4710       &SBm   &182838.9$-$665903& 313& 219& 30&1.89    &0.03   &\nodata\\
NGC~6822\d    &IBm   &194453.2$-$144811&1453&1100&330&1.21    &0.0005 &257~~  \\
NGC~6946      &SABcd &203452.0$+$600915& 818& 763&  0&0.95    &0.23   &\nodata\\
NGC~7331\d    &SAb   &223704.3$+$342435& 683& 335& 78&0.30    &0.06   &~83.3  \\
NGC~7552\d    &SAc   &231610.8$-$423505& 441& 325& 30&1.56    &0.72   &~36.3  \\
NGC~7793\d    &SAd   &235750.4$-$323530& 754& 498&  0&0.67    &0.03   &109~~  \\
\enddata     
\tablecomments{\footnotesize The ellipse parameters used in extracting optical and infrared fluxes are listed above.  The position angle is measured east of north.}
\tablecomments{\footnotesize $^\dagger$Used in the principal component analysis (see Section~\ref{sec:pca}).}
\tablenotetext{a}{\footnotesize See Section~\ref{sec:geometry}.  Entries are not included for NGC~3034 (saturated) and M81~Dwarf~A and Holmberg~IX (non-detections).}
\tablenotetext{b}{\footnotesize The equivalent radius of the ellipse including half of the total far-ultraviolet light.}
\tablenotetext{c}{\footnotesize The bright core of NGC~3034 (M~82) has rendered the {\it Spitzer} data extremely difficult to process.  Saturation effects severely limit our ability to extract reliable flux densities.}
\end{deluxetable}
\begin{deluxetable}{lcrrrrrrrrr}
\rotate
\def\a{\tablenotemark{a}}
\def\b{\tablenotemark{b}}
\def\c{\tablenotemark{c}}
\def\p{$\pm$}
\tabletypesize{\scriptsize}
\tablenum{2}
\tablecaption{Ultraviolet, Optical, and Near-Infrared Flux Densities \label{tab:uv_opt}}
\tablewidth{0pc}
\tablehead{
\colhead{Galaxy} &
\colhead{E(B-V)} &
\colhead{FUV} &
\colhead{NUV} &
\colhead{B} &
\colhead{V} &
\colhead{R} &
\colhead{I} &
\colhead{J} &
\colhead{H} &
\colhead{K$_{\rm s}$} 
\\
\colhead{} &
\colhead{} &
\colhead{1528\AA} &
\colhead{2271\AA} &
\colhead{0.45\m} &
\colhead{0.55\m} &
\colhead{0.66\m} &
\colhead{0.81\m} &
\colhead{1.25\m} &
\colhead{1.65\m} &
\colhead{2.17\m} 
\\
\colhead{} & 
\colhead{(mag)} &
\colhead{(mJy)} &
\colhead{(mJy)} &
\colhead{(Jy)} &
\colhead{(Jy)} &
\colhead{(Jy)} &
\colhead{(Jy)} &
\colhead{(Jy)} &
\colhead{(Jy)} &
\colhead{(Jy)} 
}
\startdata
NGC~0024    &0.020&  8.76~\p 1.21~&   11.43~\p 1.58~&    0.082&    0.11~&    0.11~&    0.097&    0.23~&    0.25~&    0.19~\\
NGC~0337    &0.112& 10.46~\p 1.45~&   18.69~\p 2.59~&    0.11~&    0.12~&    0.10~&    0.085&    0.20~&    0.20~&    0.17~\\
NGC~0584    &0.042&  0.37~\p 0.05~&    2.00~\p 0.28~&    0.14~&    0.28~&    0.28~&    0.29~&    0.91~&    1.12~&    0.87~\\
NGC~0628    &0.070& 75.96~\p10.52~&   99.23~\p13.74~&    0.65~&    0.84~&    0.76~&    0.65~&    1.66~&    1.67~&    1.32~\\
NGC~0855    &0.071&  1.81~\p 0.25~&    3.25~\p 0.45~&    0.034\b&  0.047\b&\nodata&  \nodata&    0.096&    0.10~&    0.085\\
NGC~0925    &0.076& 50.99~\p 7.06~&   62.43~\p 8.65~&    0.35~&    0.48~&    0.59~&    0.62~&    0.60~&    0.65~&    0.51~\\
NGC~1097    &0.027& 36.26~\p 5.19~&   50.97~\p 7.18~&    0.51~&    0.84~&    0.79~&    0.82~&    2.40~&    2.74~&    2.29~\\
NGC~1266    &0.098&  0.049\p 0.007&    0.29~\p 0.04~&    0.020&    0.036&    0.037&    0.035&    0.12~&    0.13~&    0.12~\\
NGC~1291    &0.013&  7.38~\p 1.02~&   16.28~\p 2.26~&    0.76~&    1.48~&    1.37~&    1.48~&    4.34~&    4.48~&    3.93~\\
NGC~1316    &0.021&  3.13~\p 0.44~&   16.58~\p 2.30~&    0.79~&    1.61~&    1.58~&    1.73~&    4.69~&    4.90~&    4.21~\\
NGC~1377    &0.028&  \nodata      &  \nodata        &    0.012&    0.023&    0.021&    0.033&    0.10~&    0.11~&    0.095\\
NGC~1404    &0.011&  0.97~\p 0.13~&    2.76~\p 0.38~&    0.24~&    0.48~&    0.48~&    0.49~&    1.38~&    1.59~&    1.35~\\
NGC~1482    &0.040&  0.41~\p 0.06~&    1.43~\p 0.21~&    0.024&    0.046&    0.053&    0.052&    0.23~&    0.30~&    0.29~\\
NGC~1512    &0.011& 14.95~\p 2.08~&   19.88~\p 2.77~&    0.13~&    0.25~&    0.26~&    0.21~&    0.81~&    0.86~&    0.73~\\
NGC~1566    &0.009& 54.49~\p 7.59~&   65.52~\p 9.07~&    0.43~&    0.45~&    0.47~&    0.42~&    1.39~&    1.42~&    1.27~\\
NGC~1705    &0.008& 16.01~\p 2.22~&   16.76~\p 2.32~&    0.037&    0.042&    0.036&    0.028&    0.057&    0.054&    0.044\\
NGC~2403    &0.040&258.1~~\p35.7~~&  307.5~~\p42.6~~&    1.90~&    2.42~&    2.37~&    3.45~&    2.94~&    2.91~&    2.39~\\
Holmberg~II &0.032& 47.80~\p 6.62~&   48.23~\p 6.68~&    0.21~&    0.19~&    0.25~&    0.38~&    0.22~&    0.34~&    0.26~\\
M81~Dwarf~A &0.020&  0.48~\p 0.07~&    0.56~\p 0.08~&    0.002&    0.001&    0.001&    0.002&    0.004&    0.004&    0.003\\
DDO~053     &0.038&  2.65~\p 0.37~&    2.58~\p 0.36~&    0.011&    0.008&    0.006&    0.007&    0.008&    0.014&    0.008\\
NGC~2798    &0.020&  1.12~\p 0.16~&    2.33~\p 0.32~&    0.059&    0.075&    0.071&    0.089&    0.16~&    0.19~&    0.17~\\
NGC~2841    &0.015& 12.99~\p 1.80~&   20.57~\p 2.85~&    0.85~&    1.00~&    1.26~&    1.40~&    2.81~&    3.22~&    2.67~\\
NGC~2915    &0.275& 16.13~\p 2.23~&   16.43~\p 2.27~&    0.077\b&  0.069&    0.071&    0.077&    0.13~&    0.15~&    0.092\\
Holmberg~I  &0.050&  5.29~\p 0.73~&    5.60~\p 0.78~&    0.032&    0.029&    0.015&    0.021&    0.031&    0.040&    0.016\\
NGC~2976    &0.071& 18.86~\p 2.61~&   30.24~\p 4.19~&    0.52~&    0.47~&    0.52~&    0.61~&    0.86~&    0.89~&    0.71~\\
NGC~3049\a  &0.038&  \nodata      &    4.51~\p 0.62~&    0.052&    0.051&    0.046&    0.050&    0.078&    0.082&    0.074\\
NGC~3031    &0.080&178.9~~\p24.8~~&  256.33~\p35.49~&    5.07\b&   8.73\b& \nodata&  \nodata&   23.47~&   25.44~&   21.29~\\
NGC~3034    &0.156& 50.08~\p 6.93~&  105.3~~\p14.6~~&    3.53~&    2.79\b&   3.67~&    4.74~&    9.24~&   10.80~&   10.14~\\
Holmberg~IX &0.079&  4.01~\p 0.56~&    5.00~\p 0.69~&    0.014&    0.010&    0.008&    0.010&    0.025&    0.021&    0.015\\
M81~Dwarf~B &0.081&  0.75~\p 0.10~&    0.92~\p 0.13~&    0.009&    0.007&    0.007&    0.007&    0.012&    0.014&    0.014\\
NGC~3190    &0.025&  0.40~\p 0.06~&    1.80~\p 0.25~&    0.21~&    0.27~&    0.26~&    0.37~&    0.71~&    0.84~&    0.74~\\
NGC~3184    &0.017&  \nodata      &  \nodata        &    0.67~&    0.71~&    0.70~&    1.10~&    1.05~&    1.14~&    0.91~\\
NGC~3198    &0.012& 23.60~\p 3.27~&   28.38~\p 3.93~&    0.21~&    0.30~&    0.34~&    0.42~&    0.57~&    0.63~&    0.55~\\
IC~2574     &0.036& 46.61~\p 6.45~&   48.37~\p 6.70~&    0.18~&    0.22~&    0.20~&    0.27~&    0.34~&    0.23~&    0.17~\\
NGC~3265    &0.024&  0.57~\p 0.08~&    0.96~\p 0.13~&    0.021&    0.024&    0.012&    0.024&    0.051&    0.057&    0.048\\
Markarian~33&0.012&  4.13~\p 0.57~&    5.20~\p 0.72~&    0.038&    0.034&    0.029&    0.029&    0.049&    0.056&    0.048\\
NGC~3351    &0.028& 17.66~\p 2.45~&   28.77~\p 3.98~&    0.45~&    0.58~&    0.71~&    0.98~&    1.68~&    1.77~&    1.54~\\
NGC~3521    &0.057& 22.19~\p 3.07~&   44.66~\p 6.18~&    0.89~&    1.23~&    1.40~&    2.32~&    3.73~&    4.22~&    3.50~\\
NGC~3621  &0.081& 76.91~\p11.20~&  110.2~~\p15.8~~&    0.62\b&   1.10~&  \nodata&    1.53~&    1.94~&    2.15~&    1.69~\\
NGC~3627  &0.033& 30.46~\p 4.22~&   61.43~\p 8.51~&    1.51~&    1.63~&    1.51~&    1.90~&    3.34~&    3.73~&    3.17~\\
NGC~3773  &0.027&  4.21~\p 0.58~&    5.55~\p 0.77~&    0.036&    0.033&    0.028&    0.031&    0.045&    0.039&    0.037\\
NGC~3938\a&0.021&  \nodata      &   36.41~\p 5.04~&    0.44~&    0.44~&    0.34~&    0.41~&    0.64~&    0.58~&    0.54~\\
NGC~4125\a&0.019&  \nodata      &    3.44~\p 0.48~&    0.49~&    0.54~&    0.66~&    0.87~&    1.39~&    1.54~&    1.29~\\
NGC~4236  &0.015& 63.45~\p 8.79~&   76.24~\p10.56~&    0.42~&    0.53~&    0.62~&    0.54~&    0.63~&    0.83~&    0.57~\\
NGC~4254\a&0.039&  \nodata      &   61.82~\p 8.56~&    0.78~&    0.75~&    0.64~&    0.73~&    1.27~&    1.35~&    1.21~\\
NGC~4321\a&0.026&  \nodata      &   54.04~\p 7.48~&    0.50~&    0.70~&    0.85~&    1.23~&    1.87~&    2.00~&    1.65~\\
NGC~4450\a&0.028&  \nodata      &    5.39~\p 0.75~&    0.43~&    0.53~&    0.52~&    0.65~&    1.20~&    1.39~&    1.08~\\
NGC~4536  &0.018& 16.94~\p 2.35~&   21.93~\p 3.04~&    0.40~&    0.42~&    0.47~&    0.51~&    0.71~&    0.75~&    0.70~\\
NGC~4552  &0.041&  1.89~\p 0.26~&    4.66~\p 0.65~&    0.37~&    0.49~&    0.49~&    0.58~&    1.63~&    1.80~&    1.46~\\
NGC~4559  &0.018& 53.79~\p 7.45~&   64.63~\p 8.95~&    0.66~&    0.50~&    0.50~&    0.58~&    0.77~&    0.79~&    0.66~\\
NGC~4569  &0.047&  6.00~\p 0.83~&   19.69~\p 2.73~&    0.50\b&   0.72\b& \nodata&  \nodata&    1.83~&    2.08~&    1.67~\\
NGC~4579  &0.041&  5.85~\p 0.81~&   12.11~\p 1.68~&    0.73~&    0.76~&    0.87~&    1.18~&    2.05~&    2.24~&    1.82~\\
NGC~4594  &0.051&  5.55~\p 0.77~&   17.72~\p 2.47~&    2.25~&    2.76~&    3.41~&    4.30~&    8.06~&    9.19~&    7.57~\\
NGC~4625  &0.018&  6.04~\p 0.84~&    7.97~\p 1.10~&    0.073&    0.071&    0.061&    0.071&    0.098&    0.11~&    0.089\\
NGC~4631  &0.017& 80.95~\p11.21~&  104.8~~\p14.5~~&    1.19~&    0.91~&    0.96~&    1.12~&    1.75~&    1.98~&    1.84~\\
NGC~4725  &0.012& 22.05~\p 3.07~&   29.61~\p 4.13~&    0.54~&    0.89~&    1.04~&    1.48~&    2.43~&    3.18~&    2.41~\\
NGC~4736  &0.018& 67.19~\p 9.30~&   91.87~\p12.72~&    2.50~&    2.79~&    2.76~&    3.39~&    6.94~&    7.68~&    6.44~\\
DDO~154   &0.009&  4.54~\p 0.63~&    4.42~\p 0.61~&    0.016&    0.011&    0.009&    0.009&    0.010&    0.012&    0.012\\
NGC~4826  &0.041& 14.50~\p 2.01~&   37.45~\p 5.19~&    1.41~&    2.05~&  \nodata&  \nodata&    5.67~&    6.30~&    5.28~\\
DDO~165   &0.024&  6.72~\p 0.93~&    8.15~\p 1.13~&    0.041&    0.034&    0.024&    0.023&    0.026&    0.017&    0.010\\
NGC~5033  &0.012&  \nodata      &  \nodata        &    0.54~&    0.66~&  \nodata&    0.80~&    1.21~&    1.35~&    1.17~\\
NGC~5055  &0.018& 39.30~\p 5.44~&   63.42~\p 8.78~&    1.08\b&   1.59\b& \nodata&  \nodata&    4.21~&    4.96~&    4.05~\\
NGC~5194  &0.035&160.0~~\p22.2~~&  260.8~~\p36.1~~&    1.47~&    1.96~&    2.20~&    3.02~&    4.99~&    5.89~&    4.52~\\
NGC~5195  &0.035&  3.36~\p 0.48~&   10.04~\p 1.40~&    0.37~&    0.62~&    0.81~&    1.51~&    2.37~&    2.80~&    2.26~\\
Tololo~89 &0.066&  7.57~\p 1.05~&   11.35~\p 1.57~&    0.078&    0.070&    0.050&    0.060&    0.081&    0.067&    0.054\\
NGC~5408  &0.068&  \nodata      &  \nodata        &    0.092\b&  0.11\b& \nodata&  \nodata&    0.19~&    0.17~&    0.11~\\
NGC~5474  &0.011& 24.35~\p 3.37~&   27.18~\p 3.76~&    0.13~&    0.17~&    0.18~&    0.22~&    0.14~&    0.16~&    0.11~\\
NGC~5713  &0.039&  5.16~\p 0.71~&   10.02~\p 1.39~&    0.11~&    0.14~&    0.16~&    0.20~&    0.37~&    0.39~&    0.33~\\
NGC~5866  &0.013&  0.65~\p 0.09~&    4.15~\p 0.57~&    0.48~&    0.59~&    0.60~&    0.73~&    1.31~&    1.49~&    1.26~\\
IC~4710   &0.089&  \nodata      &  \nodata        &    0.10~&    0.12~&    0.091&  \nodata&    0.11~&    0.10~&    0.078\\
NGC~6822  &0.231&306.7~~\p42.5~~&  401.9~~\p56.0~~&    1.58~&    2.24~&    1.96~&    1.49~&    5.66~&    5.64~&    4.26~\\
NGC~6946  &0.342&221.2~~\p30.8~~&  417.6~~\p58.2~~&    2.82\b&   4.10~&  \nodata&    5.08~&    7.27~&    5.47~&    5.66~\\
NGC~7331  &0.091& 15.59~\p 2.16~&   29.70~\p 4.11~&    0.54~&    0.94~&    1.09~&    1.62~&    2.85~&    3.36~&    2.82~\\
NGC~7552  &0.014&  7.73~\p 1.07~&   15.15~\p 2.11~&    0.17~&    0.26~&    0.25~&    0.23~&    0.71~&    0.80~&    0.70~\\
NGC~7793  &0.019&124.0~~\p17.2~~&  145.1~~\p20.1~~&    0.75~&    0.92~&    0.84~&    0.71~&    1.68~&    1.70~&    1.31~\\
\enddata     
\vskip-0.3in
\tablecomments{\footnotesize See \S~\ref{sec:data} for corrections that have been applied to the data.  The uncertainties include both statistical and systematic effects ($\lesssim10$\% for the optical and near-infrared data).  The 2MASS near-infrared data are from Jarret et al. (2003).}
\tablenotetext{a}{\footnotesize The far-ultraviolet detector was turned off during the observation.}
\tablenotetext{b}{\footnotesize Data from the RC3 catalog (de Vaucouleurs et al. 1991).}
\end{deluxetable}
\begin{deluxetable}{lrrrrrrrrrr}
\rotate
\def\a{\tablenotemark{a}}
\def\b{\tablenotemark{b}}
\def\c{\tablenotemark{c}}
\def\p{$\pm$}
\tabletypesize{\scriptsize}
\tablenum{3}
\tablecaption{Infrared, Submillimeter, and Radio Flux Densities \label{tab:infrared}}
\tablewidth{0pc}
\tablehead{
\colhead{Galaxy} &
\colhead{3.6\m} &
\colhead{4.5\m} &
\colhead{5.8\m} &
\colhead{8.0\m} &
\colhead{24\m} &
\colhead{70\m} &
\colhead{160\m} & 
\colhead{450\m} &
\colhead{850\m} &
\colhead{20 cm} 
\\
\colhead{} & 
\colhead{(Jy)} &
\colhead{(Jy)} &
\colhead{(Jy)} &
\colhead{(Jy)} &
\colhead{(Jy)} &
\colhead{(Jy)} &
\colhead{(Jy)} &
\colhead{(Jy)} &
\colhead{(Jy)} &
\colhead{(mJy)} 
}
\startdata
NGC~0024    & 0.10~\p 0.01~& 0.071\p 0.01~& 0.089\p 0.01~& 0.13~\p 0.02~& 0.14~\p 0.007&  2.37~\p 0.19~&   8.19~\p 1.05~ &\nodata  &\nodata   &\nodata     \\
NGC~0337    & 0.097\p 0.01~& 0.067\p 0.009& 0.14~\p 0.02~& 0.38~\p 0.05~& 0.68~\p 0.03~& 11.16~\p 0.79~&  20.09~\p 2.44~ &\nodata  &0.35\p0.05&  110  \p11 \\
NGC~0584    & 0.37~\p 0.05~& 0.22~\p 0.03~& 0.18~\p 0.02~& 0.11~\p 0.01~& 0.048\p 0.002&  0.18~\p 0.05~&   1.18~\p 0.30\a&\nodata  &\nodata   &$<$50       \\
NGC~0628    & 0.87~\p 0.12~& 0.54~\p 0.08~& 1.16~\p 0.15~& 2.70~\p 0.34~& 3.19~\p 0.13~& 34.78~\p 2.50~& 126.2~~\p15.2~~ &\nodata  &\nodata   &  173  \p17 \\
NGC~0855    & 0.043\p 0.006& 0.028\p 0.004& 0.019\p 0.003& 0.046\p 0.006& 0.087\p 0.004&  1.70~\p 0.14~&   2.50~\p 0.36~ &\nodata  &\nodata   &    4.9\p0.5\\
NGC~0925    & 0.31~\p 0.04~& 0.21~\p 0.03~& 0.35~\p 0.04~& 0.61~\p 0.08~& 0.95~\p 0.04~& 14.40~\p 1.04~&  43.33~\p 5.26~ &\nodata  &\nodata   &   46  \p~5 \\
NGC~1097    & 1.24~\p 0.17~& 0.80~\p 0.11~& 1.46~\p 0.18~& 3.19~\p 0.40~& 6.63~\p 0.27~& 59.84~\p 4.66~& 153.8~~\p18.5~~ &\nodata  &1.44\p0.78&  415  \p42 \\
NGC~1266    & 0.055\p 0.008& 0.042\p 0.006& 0.057\p 0.008& 0.090\p 0.012& 0.88~\p 0.04~& 12.69~\p 0.95~&  10.30~\p 1.29~ &\nodata  &\nodata   &  116  \p12 \\
NGC~1291    & 2.11~\p 0.29~& 1.27~\p 0.17~& 0.96~\p 0.12~& 0.64~\p 0.08~& 0.57~\p 0.02~&  6.29~\p 0.46~&  28.60~\p 3.49~ &\nodata  &\nodata   &\nodata     \\
NGC~1316    & 2.48~\p 0.34~& 1.53~\p 0.21~& 1.13~\p 0.14~& 0.55~\p 0.07~& 0.43~\p 0.02~&  5.44~\p 0.40~&  12.61~\p 1.78~ &\nodata  &\nodata   &  256  \p26 \\
NGC~1377    & 0.057\p 0.008& 0.085\p 0.012& 0.27~\p 0.04~& 0.42~\p 0.05~& 1.83~\p 0.08~&  6.35~\p 0.47~&   3.38~\p 0.42~ &\nodata  &\nodata   & $<$1.0     \\
NGC~1404    & 0.73~\p 0.10~& 0.43~\p 0.06~& 0.33~\p 0.04~& 0.16~\p 0.02~& 0.088\p 0.004&  0.17~\p 0.12\a&  0.29~\p 0.28\a&\nodata  &\nodata   &    3.9\p0.6\\
NGC~1482    & 0.21~\p 0.03~& 0.15~\p 0.02~& 0.59~\p 0.08~& 1.56~\p 0.19~& 3.69~\p 0.15~& 32.45~\p 2.88~&  38.79~\p 4.69~ &\nodata  &0.33\p0.05&  239  \p24 \\
NGC~1512    & 0.39~\p 0.05~& 0.24~\p 0.03~& 0.27~\p 0.03~& 0.44~\p 0.05~& 0.46~\p 0.02~&  6.65~\p 0.48~&  23.70~\p 2.86~ &\nodata  &\nodata   &    7.0\p~1 \\
NGC~1566    & 0.75~\p 0.10~& 0.48~\p 0.07~& 0.91~\p 0.12~& 2.11~\p 0.26~& 2.83~\p 0.13~& 34.32~\p 2.51~& 102.1~~\p12.3~~ &\nodata  &\nodata   &  400  \p40 \\
NGC~1705    & 0.026\p 0.004& 0.018\p 0.003& 0.010\p 0.002& 0.017\p 0.002& 0.056\p 0.002&  1.38~\p 0.10~&   1.66~\p 0.21~ &\nodata  &\nodata   &\nodata     \\
NGC~2403    & 1.88~\p 0.25~& 1.31~\p 0.18~& 2.13~\p 0.27~& 4.11~\p 0.51~& 5.84~\p 0.24~& 86.36~\p 6.18~& 245.6~~\p29.6~~ &\nodata  &\nodata   &  330  \p33 \\
Holmberg~II & 0.071\p 0.010& 0.057\p 0.008& 0.031\p 0.005& 0.024\p 0.005& 0.20~\p 0.008&  3.67~\p 0.26~&   4.46~\p 0.58~ &\nodata  &\nodata   &   20  \p~3 \\
M81~Dwarf~A & 0.002\p 0.001& 0.001\p 0.001& $<$0.004     & $<$0.002     & $<$0.018     &  $<$0.17      &   $<$0.15       &\nodata  &\nodata   &\nodata     \\
DDO~053     & 0.005\p 0.001& 0.004\p 0.001& 0.003\p 0.001& 0.007\p 0.001& 0.029\p 0.001&  0.40~\p 0.03~&   0.50~\p 0.11~ &\nodata  &\nodata   &\nodata     \\
NGC~2798    & 0.11~\p 0.02~& 0.081\p 0.011& 0.27~\p 0.03~& 0.63~\p 0.08~& 2.62~\p 0.11~& 21.72~\p 1.79~&  20.69~\p 2.50~ &\nodata  &0.19\p0.03&   83  \p~9 \\
NGC~2841    & 1.27~\p 0.17~& 0.75~\p 0.10~& 0.67~\p 0.09~& 1.16~\p 0.14~& 0.91~\p 0.04~& 10.22~\p 0.73~&  62.29~\p 7.54~ &\nodata  &\nodata   &   84  \p~9 \\
NGC~2915    & 0.054\p 0.008& 0.035\p 0.005& 0.033\p 0.004& 0.031\p 0.004& 0.063\p 0.003&  1.41~\p 0.11~&   1.46~\p 0.27~ &\nodata  &\nodata   &\nodata     \\
Holmberg~I  & 0.012\p 0.001& 0.008\p 0.001& 0.007\p 0.002& 0.008\p 0.002& 0.013\p 0.002&  0.42~\p 0.08~&   0.90~\p 0.17~ &\nodata  &\nodata   &\nodata     \\
NGC~2976    & 0.43~\p 0.06~& 0.28~\p 0.04~& 0.51~\p 0.07~& 1.02~\p 0.13~& 1.37~\p 0.06~& 20.43~\p 1.45~&  52.56~\p 6.35~ &\nodata  &0.61\p0.24&   51  \p~5 \\
NGC~3049    & 0.040\p 0.005& 0.028\p 0.004& 0.065\p 0.009& 0.14~\p 0.02~& 0.43~\p 0.02~&  2.90~\p 0.21~&   4.86~\p 0.59~ &\nodata  &\nodata   &   12  \p~2 \\
NGC~3031    &10.92~\p 1.48~& 6.53~\p 0.90~& 5.96~\p 0.75~& 8.04~\p 1.00~& 5.09~\p 0.20~& 85.18~\p 5.96~& 360.0~~\p43.4~~ &\nodata  &\nodata   &  380  \p38 \\
NGC~3034\b  & \nodata      & \nodata      & \nodata      & \nodata      & \nodata      &  \nodata      &   \nodata       &39.2\p9.8&5.51\p0.83& 7660~\p770 \\
Holmberg~IX & 0.007\p 0.001& 0.004\p 0.001& $<$0.013     & $<$0.012     & $<$0.037     &  $<$0.25      &   $<$0.48       &\nodata  &\nodata   &\nodata     \\
M81~Dwarf~B & 0.005\p 0.001& 0.004\p 0.001& 0.003\p 0.001& 0.003\p 0.001& 0.009\p 0.001&  0.15~\p 0.03~&   0.39~\p 0.18~ &\nodata  &\nodata   &\nodata     \\
NGC~3190    & 0.37~\p 0.05~& 0.24~\p 0.03~& 0.25~\p 0.03~& 0.33~\p 0.04~& 0.27~\p 0.01~&  5.66~\p 0.40~&  15.01~\p 1.82~ &\nodata  &0.19\p0.04&   43  \p~5 \\
NGC~3184    & 0.56~\p 0.08~& 0.36~\p 0.05~& 0.67~\p 0.08~& 1.44~\p 0.18~& 1.43~\p 0.06~& 15.76~\p 1.12~&  70.48~\p 8.50~ &\nodata  &\nodata   &   56  \p~5 \\
NGC~3198    & 0.27~\p 0.04~& 0.17~\p 0.02~& 0.34~\p 0.04~& 0.68~\p 0.09~& 1.06~\p 0.04~& 10.27~\p 0.73~&  39.00~\p 4.93~ &\nodata  &\nodata   &   27  \p~3 \\
IC~2574     & 0.15~\p 0.02~& 0.091\p 0.013& 0.066\p 0.009& 0.066\p 0.009& 0.28~\p 0.01~&  5.55~\p 0.43~&  11.75~\p 1.50~ &\nodata  &\nodata   &   11  \p~2 \\
NGC~3265    & 0.028\p 0.004& 0.020\p 0.003& 0.041\p 0.005& 0.10~\p 0.01~& 0.30~\p 0.01~&  2.71~\p 0.20~&   2.70~\p 0.34~ &\nodata  &\nodata   &   11  \p~2 \\
Markarian~33& 0.027\p 0.004& 0.019\p 0.003& 0.053\p 0.007& 0.13~\p 0.02~& 0.86~\p 0.04~&  4.35~\p 0.32~&   3.87~\p 0.48~ &\nodata  &0.04\p0.01&   17  \p~2 \\
NGC~3351    & 0.81~\p 0.11~& 0.51~\p 0.07~& 0.73~\p 0.09~& 1.33~\p 0.16~& 2.58~\p 0.12~& 24.18~\p 1.87~&  67.49~\p 8.28~ &\nodata  &\nodata   &   44  \p~5 \\
NGC~3521    & 2.05~\p 0.28~& 1.36~\p 0.19~& 2.56~\p 0.32~& 6.27~\p 0.76~& 5.51~\p 0.22~& 63.13~\p 4.54~& 222.3~~\p26.8~~ &\nodata  &2.11\p0.82&  357  \p36 \\
NGC~3621    &0.99~\p 0.13~& 0.67~\p 0.09~& 1.62~\p 0.21~&  3.51~\p 0.44~&  3.70~\p 0.19~&  50.21~\p 3.94~& 139.0~~\p17.1~~ &\nodata   &\nodata   &  198  \p20 \\
NGC~3627    &1.87~\p 0.25~& 1.25~\p 0.17~& 2.39~\p 0.30~&  5.58~\p 0.69~&  7.42~\p 0.30~&  92.63~\p 7.00~& 230.2~~\p27.7~~ &\nodata   &1.86\p0.70&  458  \p46 \\
NGC~3773    &0.022\p 0.003& 0.014\p 0.002& 0.026\p 0.004&  0.048\p 0.006&  0.14~\p 0.006&   1.58~\p 0.12~&   2.38~\p 0.33~ &\nodata   &\nodata   &    5.8\p0.5\\
NGC~3938    &0.32~\p 0.04~& 0.21~\p 0.03~& 0.41~\p 0.05~&  0.98~\p 0.12~&  1.09~\p 0.04~&  14.25~\p 1.01~&  51.98~\p 6.26~ &\nodata   &\nodata   &   62  \p~7 \\
NGC~4125    &0.64~\p 0.09~& 0.37~\p 0.05~& 0.25~\p 0.03~&  0.14~\p 0.02~&  0.079\p 0.004&   1.11~\p 0.10~&   1.77~\p 0.28~ &\nodata   &\nodata   &$<$50       \\
NGC~4236    &0.25~\p 0.03~& 0.21~\p 0.03~& 0.11~\p 0.01~&  0.22~\p 0.03~&  0.55~\p 0.02~&   8.27~\p 0.59~&  20.43~\p 2.52~ &\nodata   &\nodata   &   28  \p~3 \\
NGC~4254    &0.70~\p 0.10~& 0.47~\p 0.06~& 1.49~\p 0.19~&  3.94~\p 0.49~&  4.20~\p 0.17~&  50.29~\p 3.60~& 142.9~~\p17.2~~ &\nodata   &1.01\p0.54&  422  \p42 \\
NGC~4321    &0.95~\p 0.13~& 0.64~\p 0.09~& 1.22~\p 0.15~&  2.89~\p 0.36~&  3.34~\p 0.13~&  40.59~\p 2.90~& 139.6~~\p16.8~~ &\nodata   &0.88\p0.49&  340  \p34 \\
NGC~4450    &0.53~\p 0.07~& 0.33~\p 0.04~& 0.26~\p 0.03~&  0.27~\p 0.03~&  0.21~\p 0.01~&   3.42~\p 0.29~&  16.94~\p 2.14~ &\nodata   &\nodata   &    9.4\p~1 \\
NGC~4536    &0.40~\p 0.05~& 0.29~\p 0.04~& 0.62~\p 0.08~&  1.66~\p 0.21~&  3.46~\p 0.14~&  31.99~\p 2.49~&  58.09~\p 7.00~ &\nodata   &0.42\p0.11&  194  \p19 \\
NGC~4552    &0.83~\p 0.11~& 0.49~\p 0.07~& 0.32~\p 0.04~&  0.17~\p 0.02~&  0.094\p 0.004&   0.54~\p 0.11~&   1.42~\p 0.73~ &\nodata   &\nodata   &  100  \p~3 \\
NGC~4559    &0.35~\p 0.05~& 0.23~\p 0.03~& 0.42~\p 0.05~&  0.84~\p 0.10~&  1.12~\p 0.05~&  16.89~\p 1.20~&  54.15~\p 6.53~ &\nodata   &\nodata   &   65  \p~7 \\
NGC~4569    &0.76~\p 0.10~& 0.47~\p 0.06~& 0.59~\p 0.08~&  1.02~\p 0.13~&  1.44~\p 0.06~&  12.37~\p 0.88~&  41.21~\p 5.17~ &\nodata   &0.47\p0.08&   83  \p~9 \\
NGC~4579    &0.87~\p 0.12~& 0.52~\p 0.07~& 0.54~\p 0.07~&  0.73~\p 0.09~&  0.76~\p 0.03~&   9.53~\p 0.75~&  41.03~\p 4.95~ &\nodata   &0.44\p0.07&   98  \p10 \\
NGC~4594    &3.94~\p 0.53~& 2.31~\p 0.32~& 1.75~\p 0.22~&  1.30~\p 0.16~&  0.71~\p 0.04~&   8.02~\p 0.68~&  42.12~\p 5.58~ &\nodata   &0.37\p0.11&  137  \p14 \\
NGC~4625    &0.049\p 0.006& 0.030\p 0.004& 0.059\p 0.008&  0.13~\p 0.02~&  0.13~\p 0.006&   2.06~\p 0.16~&   5.42~\p 0.68~ &\nodata   &\nodata   &    7.1\p~2 \\
NGC~4631    &1.26~\p 0.17~& 0.84~\p 0.11~& 2.49~\p 0.31~&  5.86~\p 0.73~&  8.15~\p 0.33~& 130.2~~\p 9.9~~& 289.5~~\p34.9~~ &30.7\p10.0&5.73\p1.21& 1200~~\p120\\
NGC~4725    &1.14~\p 0.15~& 0.70~\p 0.10~& 0.75~\p 0.10~&  1.21~\p 0.15~&  0.86~\p 0.04~&   8.85~\p 0.66~&  59.91~\p 7.36~ &\nodata   &\nodata   &   28  \p~3 \\
NGC~4736    &3.60~\p 0.49~& 2.32~\p 0.32~& 2.76~\p 0.35~&  5.17~\p 0.64~&  5.65~\p 0.23~&  93.93~\p 7.34~& 177.4~~\p21.4~~ &\nodata   &1.54\p0.66&  271  \p27 \\
DDO~154     &0.004\p 0.001& 0.003\p 0.001&$<$0.006      &  $<$0.004     &  0.008\p 0.001\a& 0.065\p 0.05\a&  0.35~\p 0.12\a&\nodata   &\nodata   &\nodata     \\
NGC~4826    &2.52~\p 0.34~& 1.57~\p 0.22~& 1.66~\p 0.21~&  2.35~\p 0.29~&  2.72~\p 0.15~&  55.16~\p 5.05~&  98.82~\p12.67~ &\nodata   &1.23\p0.31&  101  \p10 \\
DDO~165     &0.016\p 0.002& 0.012\p 0.002& 0.005\p 0.002&  0.004\p 0.001\a&0.014\p 0.001\a& 0.15~\p 0.07\a&  0.33~\p 0.26\a&\nodata   &\nodata   &\nodata     \\
NGC~5033    &0.64~\p 0.09~& 0.47~\p 0.06~& 0.82~\p 0.10~&  1.92~\p 0.24~&  1.97~\p 0.08~&  28.81~\p 2.09~&  91.07~\p11.2~~ &\nodata   &1.10\p0.55&  178  \p18 \\
NGC~5055    &2.38~\p 0.32~& 1.55~\p 0.21~& 2.67~\p 0.34~&  5.64~\p 0.70~&  5.73~\p 0.23~&  72.57~\p 5.16~& 302.3~~\p36.6~~ &\nodata   &\nodata   &  390  \p39 \\
NGC~5194    &2.66~\p 0.36~& 1.80~\p 0.25~& 4.29~\p 0.54~& 10.64~\p 1.32~& 12.67~\p 0.53~& 147.1~~\p10.6~~& 494.7~~\p59.8~~ &\nodata   &2.61\p0.39& 1490~~\p150\\
NGC~5195    &0.83~\p 0.11~& 0.51~\p 0.07~& 0.47~\p 0.06~&  0.65~\p 0.08~&  1.40~\p 0.27~&  16.31~\p 1.89~&  14.86~\p 2.52~ &\nodata   &0.26\p0.04&   50  \p~5 \\
Tololo~89   &0.038\p 0.005& 0.025\p 0.004& 0.014\p 0.002&  0.059\p 0.008&  0.28~\p 0.01~&   2.03~\p 0.16~&   3.52~\p 0.51~ &\nodata   &\nodata   &    4.2\p0.8\\
NGC~5408    &0.052\p 0.007& 0.037\p 0.005& 0.041\p 0.005&  0.038\p 0.005&  0.43~\p 0.02~&   3.59~\p 0.27~&   2.57~\p 0.38~ &\nodata   &\nodata   &\nodata     \\
NGC~5474    &0.10~\p 0.01~& 0.073\p 0.010& 0.077\p 0.010&  0.12~\p 0.01~&  0.18~\p 0.008&   3.73~\p 0.27~&  10.56~\p 1.29~ &\nodata   &\nodata   &   12  \p~2 \\
NGC~5713    &0.20~\p 0.03~& 0.14~\p 0.02~& 0.30~\p 0.04~&  1.16~\p 0.15~&  2.35~\p 0.10~&  23.69~\p 1.84~&  39.66~\p 4.79~ &\nodata   &0.57\p0.12&  160  \p16 \\
NGC~5866    &0.66~\p 0.09~& 0.42~\p 0.06~& 0.31~\p 0.04~&  0.31~\p 0.04~&  0.21~\p 0.009&   8.71~\p 0.63~&  17.74~\p 2.14~ & 0.8\p 0.2&0.14\p0.02&   23  \p~3 \\
IC~4710     &0.070\p 0.010& 0.047\p 0.007& 0.045\p 0.006&  0.065\p 0.008&  0.12~\p 0.005&   2.37~\p 0.18~&   3.57~\p 0.48~ &\nodata   &\nodata   &\nodata     \\
NGC~6822    &2.12~\p 0.29~& 1.38~\p 0.19~& 1.45~\p 0.18~&  1.41~\p 0.18~&  3.18~\p 0.13~&  63.75~\p 4.50~& 143.5~~\p17.4~~ &\nodata   &\nodata   &   69  \p14 \\
NGC~6946    &3.31~\p 0.45~& 2.18~\p 0.30~& 5.88~\p 0.74~& 14.12~\p 1.76~& 20.37~\p 0.81~& 207.2~~\p16.1~~& 502.8~~\p60.6~~ &18.5\p 4.6&2.98\p0.45& 1395~~\p140\\
NGC~7331    &1.61~\p 0.22~& 1.02~\p 0.14~& 1.87~\p 0.24~&  4.05~\p 0.50~&  4.36~\p 0.25~&  74.97~\p 6.62~& 189.5~~\p24.3~~ &20.6\p 8.1&2.11\p0.38&  373  \p37 \\
NGC~7552    &0.45~\p 0.06~& 0.36~\p 0.05~& 1.07~\p 0.14~&  2.71~\p 0.34~& 10.66~\p 0.44\c& 67.59~\p11.1~\c& 93.39~\p11.25~ &\nodata   &0.80\p0.17&  276  \p28 \\
NGC~7793    &0.77~\p 0.10~& 0.47~\p 0.06~& 1.04~\p 0.13~&  1.85~\p 0.23~&  2.05~\p 0.08~&  34.29~\p 2.43~& 126.2~~\p15.3~~ &\nodata   &\nodata   &  103  \p10 \\
\enddata
\tablecomments{\footnotesize See \S~\ref{sec:data} for details on the data.  Upper limits (3$\sigma$) are provided for non-detections.}
\tablenotetext{a}{\footnotesize Possibly severely contaminated by background source(s).}
\tablenotetext{b}{\footnotesize The bright core of NGC~3034 (M~82) has rendered the {\it Spitzer} data extremely difficult to process.  Saturation effects severely limit our ability to extract reliable global flux densities.}
\tablenotetext{c}{\footnotesize Flux artificially low due to saturation effects.}
\end{deluxetable}
\begin{deluxetable}{cccc}
\tabletypesize{\scriptsize}
\tablenum{4}
\tablecaption{IRAC Aperture Correction Parameters \label{tab:IRAC_ap}}
\tablewidth{0pc}
\tablehead{
\colhead{$\lambda$} &
\colhead{A} &
\colhead{B} &
\colhead{C} 
}
\startdata
3.5\m&   0.82&    0.370&   0.910\\
4.5\m&   1.00&    0.380&   0.940\\
5.8\m&   1.49&    0.207&   0.720\\
8.0\m&   1.37&    0.330&   0.740\\
\enddata
\tablecomments{\footnotesize See \S~\ref{sec:data} and spider.ipac.caltech.edu/staff/jarrett/irac/}
\end{deluxetable}
\begin{deluxetable}{lccccc}
\def\a{\tablenotemark{a}}
\def\b{\tablenotemark{b}}
\def\c{\tablenotemark{c}}
\def\p{$\pm$}
\tabletypesize{\scriptsize}
\tablenum{5}
\tablecaption{Infrared and Submillimeter Aperture Correction Factors \label{tab:submm_radio}}
\tablewidth{0pc}
\tablehead{
\colhead{Galaxy} &
\colhead{24\m} &
\colhead{70\m} &
\colhead{160\m} &
\colhead{450\m} &
\colhead{850\m} 
}
\startdata
NGC~0024     &1.06&1.10&1.20&\nodata&\nodata\\
NGC~0337     &1.01&1.06&1.15&\nodata&\nodata\\
NGC~0584     &1.00&1.04&1.11&\nodata&\nodata\\
NGC~0628     &1.02&1.03&1.06&\nodata&\nodata\\
NGC~0855     &1.02&1.07&1.15&\nodata&\nodata\\
NGC~0925     &1.03&1.04&1.07&\nodata&\nodata\\
NGC~1097     &1.01&1.02&1.06&\nodata&   2.09\\
NGC~1266     &1.01&1.05&1.13&\nodata&\nodata\\
NGC~1291     &1.01&1.02&1.04&\nodata&\nodata\\
NGC~1316     &1.06&1.02&1.18&\nodata&\nodata\\
NGC~1377     &1.02&1.06&1.15&\nodata&\nodata\\
NGC~1404     &1.00&1.02&1.07&\nodata&\nodata\\
NGC~1482     &1.00&1.03&1.10&\nodata&\nodata\\
NGC~1512     &1.04&1.05&1.06&\nodata&\nodata\\
NGC~1566     &1.04&1.05&1.04&\nodata&\nodata\\
NGC~1705     &1.04&1.12&1.19&\nodata&\nodata\\
NGC~2403     &1.00&1.00&1.02&\nodata&\nodata\\
Holmberg~II  &1.00&1.02&1.09&\nodata&\nodata\\
M81~Dwarf~A  &1.00&1.00&1.00&\nodata&\nodata\\
DDO~053      &1.03&1.15&1.48&\nodata&\nodata\\
NGC~2798     &1.03&1.08&1.19&\nodata&   1.08\\
NGC~2841     &1.01&1.04&1.10&\nodata&\nodata\\
NGC~2915     &1.07&1.15&1.33&\nodata&\nodata\\
Holmberg~I   &1.01&1.05&1.15&\nodata&\nodata\\
NGC~2976     &1.00&1.03&1.10&\nodata&   1.56\\
NGC~3049     &1.02&1.07&1.18&\nodata&\nodata\\
NGC~3031     &1.00&1.00&1.01&\nodata&\nodata\\
NGC~3034     &1.00&1.00&1.00&\nodata&\nodata\\
Holmberg~IX  &1.00&1.00&1.00&\nodata&\nodata\\
M81~Dwarf~B  &1.12&1.16&1.86&\nodata&\nodata\\
NGC~3190     &1.01&1.05&1.13&\nodata&   1.12\\
NGC~3184     &1.00&1.00&1.05&\nodata&\nodata\\
NGC~3198     &1.00&1.02&1.08&\nodata&\nodata\\
IC~2574      &1.04&1.07&1.12&\nodata&\nodata\\
NGC~3265     &1.02&1.07&1.16&\nodata&\nodata\\
Markarian~33 &1.02&1.07&1.16&\nodata&\nodata\\
NGC~3351     &1.04&1.06&1.11&\nodata&\nodata\\
NGC~3521     &1.00&1.01&1.05&\nodata&   1.56\\
NGC~3621     &1.06&1.07&1.09&\nodata&\nodata\\
NGC~3627     &1.00&1.01&1.06&\nodata&   1.53\\
NGC~3773     &1.07&1.13&1.15&\nodata&\nodata\\
NGC~3938     &1.01&1.03&1.08&\nodata&\nodata\\
NGC~4125     &1.09&1.16&1.31&\nodata&\nodata\\
NGC~4236     &1.00&1.02&1.06&\nodata&\nodata\\
NGC~4254     &1.00&1.02&1.07&\nodata&   2.06\\
NGC~4321     &1.00&1.01&1.06&\nodata&   2.19\\
NGC~4450     &1.04&1.08&1.16&\nodata&\nodata\\
NGC~4536     &1.00&1.02&1.08&\nodata&   1.30\\
NGC~4552     &1.00&1.03&1.11&\nodata&\nodata\\
NGC~4559     &1.00&1.03&1.09&\nodata&\nodata\\
NGC~4569     &1.00&1.01&1.07&\nodata&   1.11\\
NGC~4579     &1.01&1.04&1.07&\nodata&\nodata\\
NGC~4594     &1.06&1.09&1.17&\nodata&   1.33\\
NGC~4625     &1.01&1.06&1.16&\nodata&\nodata\\
NGC~4631     &1.00&1.00&1.05&   1.27&   1.17\\
NGC~4725     &1.03&1.05&1.10&\nodata&\nodata\\
NGC~4736     &1.00&1.00&1.02&\nodata&   1.67\\
DDO~154      &1.06&1.14&1.35&\nodata&\nodata\\
NGC~4826     &1.08&1.09&1.14&\nodata&   1.24\\
DDO~165      &1.02&1.09&1.24&\nodata&\nodata\\
NGC~5033     &1.00&1.01&1.05&\nodata&   1.93\\
NGC~5055     &1.00&1.00&1.03&\nodata&\nodata\\
NGC~5194     &1.00&1.00&1.01&\nodata&\nodata\\
NGC~5195     &1.01&1.06&1.13&\nodata&\nodata\\
Tololo~89    &1.08&1.14&1.22&\nodata&\nodata\\
NGC~5408     &1.01&1.05&1.12&\nodata&\nodata\\
NGC~5474     &1.00&1.03&1.09&\nodata&\nodata\\
NGC~5713     &1.01&1.06&1.14&\nodata&   1.17\\
NGC~5866     &1.00&1.03&1.09&\nodata&\nodata\\
IC~4710      &1.03&1.08&1.19&\nodata&\nodata\\
NGC~6822     &1.00&1.00&1.01&\nodata&\nodata\\
NGC~6946     &1.00&1.00&1.03&\nodata&\nodata\\
NGC~7331     &1.08&1.10&1.16&   1.44&   1.11\\
NGC~7552     &1.01&1.02&1.13&\nodata&   1.17\\
NGC~7793     &1.01&1.03&1.08&\nodata&\nodata\\
\enddata     
\tablecomments{\footnotesize IRAC aperture corrections are described by Equation~\ref{eqn:irac}.  See \S~\ref{sec:infrared} and Dale et al. (2005) for details.}
\end{deluxetable}

\clearpage

\clearpage
\begin{figure}
 \caption{Examples of galaxies with clumpy (Holmberg~II), unresolved (NGC~1266), and smooth (NGC~2841) 24\m\ emission.  The left, middle, and right panels respectively show the original 24\m\ images, images of the point sources therein, and the differences in the original and point source images (see \S~\ref{sec:geometry}).  The images are approximately 700\arcsec\ across ($\sim$12~kpc, $\sim$100~kpc, and $\sim$35~kpc for Holmberg~II, NGC~1266, and NGC~2841).}
 \label{fig:mosaic}
\end{figure}

\clearpage

\clearpage

\end{document}